\pdfoutput=1
\documentclass[twocolumn,showpacs,prb,amsmath,amssymb,floatfix,eqsecnum,10pt,aps]{revtex4-1}

\usepackage{amsmath,amssymb}
\usepackage{bm}
\usepackage{graphicx}
\usepackage{psfrag}
\usepackage{color}

\newcommand{\bd}{\bm}

\newcommand{\Ref}[1]{
\!\!Eq.~(\ref{#1})}
\newcommand{\vt}{\theta}
\begin{document}

\title{Elastic constants and ultrasonic attenuation in the cone state of
the frustrated antiferromagnet Cs$_2$CuCl$_4$
}

\author{Andreas Kreisel and Peter Kopietz}
  
\affiliation{Institut f\"{u}r Theoretische Physik, Universit\"{a}t
  Frankfurt,  Max-von-Laue Strasse 1, 60438 Frankfurt, Germany}

\author{{Pham Thanh} Cong, Bernd Wolf, and Michael Lang}
  
\affiliation{Physikalisches Institut, Universit\"{a}t
  Frankfurt,  Max-von-Laue Strasse 1, 60438 Frankfurt, Germany}

\date{July 15, 2011}

 \begin{abstract}

In an external magnetic field perpendicular to the plane of the
layers, the quasi two-dimensional frustrated antiferromagnet
 Cs$_2$CuCl$_4$  exhibits a magnetically ordered ``cone state'' 
at low temperatures. In this state the component of
the magnetic moments in field direction is finite, while
their projection onto the plane of the layers forms a spiral.
We present  both theoretical and experimental
results for the magnetic-field dependence of the
elastic constants and the ultrasonic attenuation rate in the cone state.
Our theoretical analysis is based on the usual spin-wave expansion around the
classical ground state of a
Heisenberg model on an anisotropic triangular lattice with Dzyaloshinskii--Moriya
interactions. Magnon-phonon interactions are modeled by expanding
the exchange interactions up to second order in powers of the phonon 
coordinates.  As long as the external magnetic field is not too close to the
critical field where the cone state becomes unstable, we obtain reasonable agreement 
between theory and experiment, suggesting that at least in this regime magnons 
behave as well-defined quasiparticles.
We also show that the assumption of well-defined magnons 
implies that at long wavelengths the ultrasonic attenuation rate in the cone state of
 Cs$_2$CuCl$_4$ is proportional to the fourth power of the
phonon momentum.

\end{abstract}


\maketitle

\section{Introduction}

In recent years a large amount of 
theoretical and experimental work has been devoted to  
low-dimensional quantum magnets on frustrated lattices.\cite{*[{For a collection of
recent reviews see, 
for example, }][]Lacroix10} 
In these systems enhanced quantum fluctuations and competing interactions
can stabilize magnetically disordered ground states in certain parameter regimes.
The phase transitions between ordered and disordered phases as a function of
some nonthermal control parameter, such as an external magnetic field,
have attracted a lot of attention. 
An important realization of such a frustrated magnet 
exhibiting a rather complex phase diagram, including magnetically ordered and
disordered phases, is
the magnetic insulator Cs$_2$CuCl$_4$.
This material 
was synthesized and characterized about 15 years ago.\cite{Nasyrov94, Coldea96} 
Coldea~\textit{et~al.}\cite{Coldea97} determined the magnetic structure and pointed out that 
it is a realization of a triangular lattice antiferromagnet with two 
different  exchange couplings $J$ and $J^{\prime}$ whose magnitude is less than $1~\mathrm{meV}$.
The effective $g$-factor was obtained by electron spin resonance measurements\cite{Bailleul94}.
Later on, an effective model Hamiltonian was deduced from high field measurements.\cite{Coldea02}
The observation of Bose-Einstein condensation of magnons\cite{Radu05} and 
investigations on the phase diagram followed. 
An ordered cone state and a ferromagnetic phase which are connected by a quantum critical point were discovered.
The challenge to gain a quantitative microscopic understanding of the
experimental observations has  motivated a large number of recent theoretical 
works.\cite{Veillette05a,Veillette05,Veillette06,Dalidovich06,Kovrizhin06,Weng06,Kohno07,Starykh07,Kohno09,Starykh10}

There are two complementary theoretical points of view:
On the one hand, it has been argued that the difference in the two competing exchange
couplings in the layers is sufficiently large to view the system as a collection of weakly
coupled spin chains.\cite{Starykh07,Kohno07,Starykh07,Kohno09,Starykh10}
The properties  of the decoupled one-dimensional spin chains can then be
obtained nonperturbatively using bosonization techniques, while the weak coupling
between the chains is usually included using some kind of mean-field approximation.
Because in this approach one assumes that the elementary excitations 
of the system can be connected to the spinons of the one-dimensional 
spin chains, this approach 
is most natural to describe the spin-liquid sector of the phase diagram.
On the other hand, for the description of the part of the phase diagram 
which exhibits magnetic order, it seems to be more natural 
to start from the usual spin-wave picture and expand in fluctuations around
an ordered classical ground state using the 
Holstein--Primakoff 
transformation.\cite{Chung01,Dalidovich06,Kovrizhin06,Veillette05a,Veillette05,Veillette06,Fjaerestad07} Formally, this approach is based on the smallness of the inverse spin quantum number
$1/S$. However, even for $S=1/2$ the result of the $1/S$-expansion has often been surprisingly
accurate, at least for quantum antiferromagnets on nonfrustrated lattices.
It is therefore reasonable to expect that the $1/S$-expansion is a good starting point 
to describe the magnetically ordered
cone state of Cs$_2$CuCl$_4$, which is stable
at sufficiently low temperatures and in a range of magnetic fields 
oriented perpendicular to the layers. 
It turns out, however, that the $1/S$-expansion in the cone state
has been only partially successful because experimentally observed large quantum
renormalizations of the exchange energies could not be explained 
by including only the leading $1/S$ corrections to linear spin-wave theory.
In particular, Veillette~\textit{et~al.}\cite{Veillette05}
calculated the dynamical structure factor in the cone state of Cs$_2$CuCl$_4$,
including the leading $1/S$ corrections to linear spin-wave theory.
Although they found that the  spin-wave interactions 
explain, on a qualitative level, many experimentally observed features, 
there was no  quantitative agreement between theory and experiment.
Whether or not infinite resummations of the $1/S$-expansion
retaining certain types of interaction processes 
would improve the agreement with experiments remains  an open problem.

The purpose of this paper is to
further investigate the validity of the spin-wave picture
in the cone state of Cs$_2$CuCl$_4$. Therefore, we present both experimental and
theoretical results for the magnetic-field dependence of the elastic constants and the
ultrasonic attenuation rate in Cs$_2$CuCl$_4$
at low temperatures and in the range of magnetic fields where the cone state
is stable.
Since the lattice vibrations are coupled to the spin excitations, 
the magnetic-field dependence of the energy and the damping of the phonon excitations
provides useful information about the magnetic excitations.
Indeed, in the vicinity of a magnetic phase transition one expects rather strong
magnetic-field dependence of the phonon degrees of freedom, so that
one can map out the phase diagram by measuring the elastic constants and the
phonon damping 
which determines the ultrasonic attenuation rate.~\cite{Luethi}

Previous ultrasonic investigations of the longitudinal $c_{22}$-mode on single crystals of Cs$_2$CuCl$_4$ 
by Sytcheva {\it{et~al.}}~\cite{Sytcheva09} focused
on the nature of the different phases and on the
field-induced quantum critical point.
Our theoretical approach is complementary to the
approach adopted by the authors of Ref.~[\onlinecite{Sytcheva09}], who
obtained the elastic constants from simple phenomenological
expressions for the thermodynamic potentials~\cite{Luethi}
and relied on phenomenological  relaxation rates
to determine the ultrasonic attenuation rate.~\cite{Sytcheva09}
In contrast, our derivation of the ultrasonic attenuation rate and the elastic constants is 
based on a microscopic calculation starting from
the relevant Heisenberg Hamiltonian. 
Although  the theory
of magnetoelastic interactions is well established~\cite{Luethi,Tachiki74} 
there have been only a few previous investigations of the interplay
between spins and lattice vibrations in triangular lattice 
antiferromagnets\cite{Lim04,Aplesnin04,Kim07} and other
frustrated spin systems.\cite{Zhou11}
In particular, microscopic investigations of the
interaction between spin  and lattice degrees of freedom
in the cone state of  Cs$_2$CuCl$_4$ have so far not been carried out.

The rest of this paper is organized as follows: 
Our starting point in
Sec.~\ref{sec:spinpho} is the assumption that the
interplay between magnetic and lattice degrees of freedom
in Cs$_2$CuCl$_4$ can be described by a spin $S=1/2$
quantum Heisenberg antiferromagnet on a distorted triangular lattice
with Dzyaloshinskii--Moriya (DM) anisotropy in an external magnetic field.
The coupling between magnons and phonons arises from the fact that the 
exchange integrals and the DM interactions depend on the distances between the actual
positions of the ions, which in turn depend on the phonon coordinates.
Assuming an ordered classical ground state (the cone state discussed in 
Sec.~\ref{sec:spinwave}), we model the  low-lying spin excitations 
in terms of suitably defined interacting Holstein--Primakoff bosons.
By expanding the exchange integrals to second order in the phonon coordinates
we obtain in Sec.~\ref{sec:magpho} the effective magnon-phonon interaction.
In particular, we explicitly derive the  relevant interaction vertices, which exhibit rather
complicated momentum dependencies due to the spiral spin structure
of the cone state.

In Sec.~\ref{sec:renormalization} we then use functional integration techniques 
and diagrammatic perturbation theory to calculate the renormalization of the
phonon energies due to the coupling to the magnons. 
Formally, this renormalization can be described in terms of a
momentum- and frequency-dependent phonon self-energy, whose real part
gives the shift in the phonon velocities (which are related to the elastic constants),
and whose imaginary part gives the phonon damping (which is related to the
ultrasonic attenuation rate).
Using the fact that in Cs$_2$CuCl$_4$  the velocities of acoustic phonons are large
compared with the magnon velocities, we can derive analytic expressions for
the magnetic-field-dependent part of the elastic constants and the ultrasonic attenuation rate.
In particular, we show that the for small 
wavevectors $\bd{k}$ the ultrasonic attenuation rate
 in Cs$_2$CuCl$_4$ is proportional to $\bd{k}^4$. 

In Sec.~\ref{sec:comparison} we present our experimental results for 
the magnetic-field dependence of the elastic constants 
and the ultrasonic attenuation rate in the cone state of
Cs$_2$CuCl$_4$  and compare them with our theoretical predictions.
In the regime of magnetic fields where our perturbative calculation is valid
we find good agreement between theory and experiment.
Finally, in Sec.~\ref{sec:summary} we present our conclusions and discuss some
open problems.  In the appendix we evaluate the contribution of scattering processes 
involving intermediate states with one phonon and one magnon 
to the ultrasonic attenuation rate; we show that in
Cs$_2$CuCl$_4$ the smallness of the
magnon velocities in comparison with the phonon velocities
guarantees that these processes are small  compared with the
processes considered in Sec.~\ref{subsec:ultrasound}.

\section{Magnon-phonon interactions in the cone state of $\mathrm{Cs}_2\mathrm{CuCl}_4$
}
\label{sec:hamiltonian}
\subsection{Spin-phonon Hamiltonian}
\label{sec:spinpho}

To model the magnetoelastic properties of the frustrated quantum antiferromagnet
Cs$_2$CuCl$_4$, we start from the following
spin-phonon Hamiltonian,
 \begin{equation}
 {H} =      {H}^{\mathrm{pho}}_{\mathrm{ spin}} +  {H}^{\mathrm{pho}}    ,
\label{eq:Hs0}
 \end{equation}
where the first term  is of the form~\cite{Coldea02,Veillette05a,Veillette05} 
\begin{align}
 {H}^{\mathrm{pho}}_{\mathrm {spin}}   & = \frac{1}{2} \sum_{ i j  } 
 \left[ 
J_{ij} {\bd{S}}_i \cdot {\bd{S}}_j +
 \bd{D}_{ ij} \cdot (  {\bd{S}}_i \times {\bd{S}}_j ) \right]
  -      \sum_{i} \bd{h} \cdot {\bd{S}}_i.
 \label{eq:Hs}
 \end{align}
Here the ${\bd{S}}_i$ are spin $S=1/2$ operators at positions
$\bd{r}_i$,  the $J_{ij}$ are the exchange energies connecting
spins at positions ${\bd{r}}_i$ and $\bd{r}_j$, 
the antisymmetric vectors $\bd{D}_{ ij}$ model the DM interaction, and $\bd{h}$ is an external magnetic field.
The second term in Eq.~(\ref{eq:Hs0})  
describes noninteracting acoustic phonons, 
 \begin{equation}
{H}^{ \mathrm{pho}} =   \sum_{ \bd{k}  \lambda } \omega_{\bd{k} \lambda } \left( a^{\dagger}_{ \bd{k} \lambda}
 a_{\bd{k} \lambda} + \frac{1}{2} \right),
 \label{eq:Hp}
 \end{equation}
where $a_{\bd{k} \lambda}$ annihilates a phonon with wavevector $\bd{k}$ and
polarization $\lambda$, and the long-wavelength dispersion of the phonons is
\begin{equation}
\omega_{\bd{k} \lambda} = c_{\lambda} ( \hat{\bd{k}} ) | \bd{k} |.
\label{eq:defspectrph}
 \end{equation}
The velocities $c_{\lambda} ( \hat{\bd{k}} )$ of the phonons depend on their 
propagation direction $\hat{\bd{k}} = \bd{k}  / | \bd{k} |$.
The spin-phonon interaction in Eq.~(\ref{eq:Hs})
arises from the fact that
 $\bd{r}_i$ and $\bd{r}_j$ are the actual positions of the spins
for a given configuration of lattice distortions $\{ \bd{X}_i \}$.
Denoting by $\{ \bd{R}_i \}$ the positions of the
corresponding Bravais lattice, we have ${\bd{r}}_i = \bd{R}_i + \bd{X}_i$.
The lattice distortions are quantized in the usual way,
 \begin{subequations}
 \begin{align}
 \bd{X}_i & =  \frac{1}{\sqrt{N}} \sum_{\bd{k}} e^{ i \bd{k} \cdot \bd{R}_i } 
 \bd{X}_{\bd{k}},
 \\
 \bd{X}_{\bd{k}} & =  \sum_{\lambda} X_{ \bd{k} \lambda } \bd{e}_{ \bd{k} \lambda } ,
 \\
  X_{ \bd{k} \lambda } & =   \frac{1}{ \sqrt{2 M \omega_{\bd{k} \lambda }} }
 ( a_{ \bd{k} \lambda } +  a_{ -\bd{k} \lambda }^{\dagger} ),
 \end{align}
\end{subequations}
where the unit vectors $\bd{e}_{\bd{k} \lambda}$ define the polarization
directions of the phonons.
Introducing the momentum operators conjugate to the $X_{\bd{k} \lambda}$,
 \begin{equation}
 P_{ \bd{k} \lambda }  =   \frac{1}{i} \sqrt{ \frac{  M \omega_{\bd{k} \lambda }}{2}  }
 ( a_{ \bd{k} \lambda } -  a_{ -\bd{k} \lambda }^{\dagger} ),
\end{equation}
the pure phonon Hamiltonian (\ref{eq:Hp}) can be written as
 \begin{equation}
{H}^{\mathrm{ pho}} =    \sum_{ \bd{k}  \lambda } \left[
 \frac{ P_{ - \bd{k} \lambda} P_{  \bd{k} \lambda}}{2 M} +
 \frac{M}{2} \omega^2_{ \bd{k} \lambda}  X_{ - \bd{k} \lambda} X_{  \bd{k} \lambda}
 \right].
 \end{equation}
Assuming that positions of the spins deviate only slightly from their equilibrium values,
we may expand the exchange couplings  $J_{ ij}$ and the DM
vectors $\bd{D}_{ ij}$ in powers of the
difference vectors $\bd{X}_{ ij} = \bd{X}_i - \bd{X}_j$,
 \begin{subequations}
 \begin{align}
 J_{ ij }  & =  J ( \bd{R}_{ij} ) + ( \bd{X}_{ ij} \cdot \mathbf{\nabla}_{\bd{r}} )
 \left. J ( \bd{r} ) \right|_{ \bd{r} = \bd{R}_{ ij} } 
 \nonumber
 \\
 & + 
\frac{1}{2} ( \bd{X}_{ ij} \cdot \mathbf{\nabla}_{\bd{r}} )^2
 \left. J ( \bd{r} ) \right|_{ \bd{r} = \bd{R}_{ ij} } +
\cdots ,
 \label{eq:Jgrad}
 \\
{\bd{D}}_{ ij }  & =  {\bd{D}} ( \bd{R}_{ij} ) + ( \bd{X}_{ ij} \cdot \mathbf{\nabla}_{\bd{r}} )
 \left. {\bd{D}} ( \bd{r} ) \right|_{ \bd{r} = \bd{R}_{ ij} } 
 \nonumber
 \\
 & 
+
\frac{1}{2} ( \bd{X}_{ ij} \cdot \mathbf{\nabla}_{\bd{r}} )^2
\left. {\bd{D}} ( \bd{r} ) \right|_{ \bd{r} = \bd{R}_{ ij} } 
+ \cdots ,
 \label{eq:Dgrad}
 \end{align}
 \end{subequations}
where $\bd{R}_{ij} = \bd{R}_i - \bd{R}_j$ is again a vector of the Bravais lattice.
Substituting the expansions (\ref{eq:Jgrad}, \ref{eq:Dgrad})
into Eq.~(\ref{eq:Hs}) we obtain an expansion of
our spin-phonon
Hamiltonian in powers of the phonon operators,
 \begin{equation}
 H^{\mathrm{pho}}_{\mathrm{spin}} = H_{\mathrm{spin}} + H^{\mathrm{1pho}}_{\mathrm{spin}} + H^{\mathrm{2pho}}_{\mathrm{spin}} + \cdots,
 \label{eq:Hsp}
 \end{equation}
where the pure spin part $H_{\mathrm{spin}}$ is obtained by assuming that the spins 
are located at the
sites $\bd{R}_i$ of the Bravais lattice, while the spin-phonon interactions $H^{\mathrm{1pho}}_{\mathrm{spin}}$ 
and $H^{\mathrm{2pho}}_{\mathrm{spin}}$
describe the
coupling of one and two powers of the phonon 
operators $\bd{X}_{ ij}$
 to  the spin system.

\begin{figure}[tb]
  \includegraphics[width=0.45\textwidth]{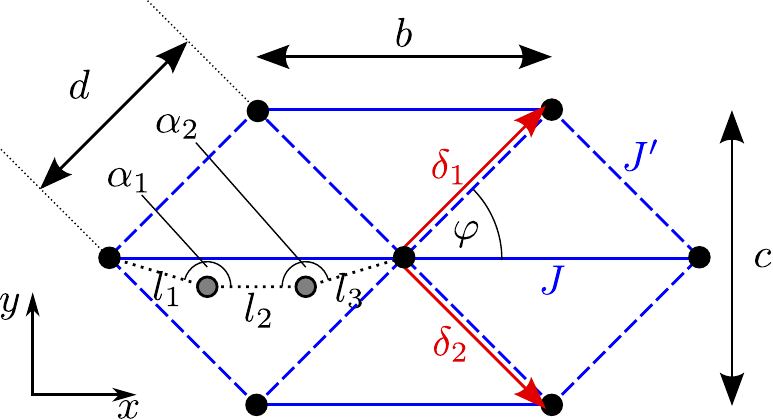}
 \caption{(Color online) Triangular lattice with the lattice parameters $b$ and $c$ describing the geometry
of a single layer of Cs$_2$CuCl$_4$.
The magnetic ions are located at the edges of the triangles (black dots) and are mainly coupled via the coupling $J$ (solid line) in $x$-direction and $J'$ along the diagonals (dashed line). The gray dots indicate the position of the chlorine atoms mediating the exchange 
interaction, 
which depends in a simple model on the three distances $l_1$, $l_2$, $l_3$ 
and on the two bonding angles $\alpha_1$ and $\alpha_2$.  
For simplicity, we lump these
dependencies into the variation of the bond length. In the same manner, 
also the dependence of the coupling $J'$ on the angle $\varphi$ is neglected.}
 \label{fig:lattice}
 \end{figure}

The pure spin part $H_{\mathrm{spin}}$
of our spin-phonon Hamiltonian (\ref{eq:Hsp}) which is formally
identical with Eq.~(\ref{eq:Hs})  except that now the spins are located at the
sites $\bd{R}_i$ of the Bravais lattice.
From the values compiled in Tab.~\ref{tab:cscucl}
it is clear that for the layered material Cs$_2$CuCl$_4$ the couplings between
neighboring layers is very small~\cite{Coldea02,Coldea03} so that we may ignore
the interlayer coupling $J^{\prime \prime}$ and focus on a single layer.
The spins are then 
located at the sites of a distorted triangular  lattice
characterized by the lattice parameters $b$ and $c$, as shown
in Fig.~\ref{fig:lattice}.
Within the planes of the lattice,
we assume an anisotropic nearest-neighbor exchange,~\cite{Coldea02,Veillette05a,Veillette05} 
\begin{equation}
 J_{ ij} = J ( \bd{R}_i - \bd{R}_j ) = \left\{
 \begin{array}{ll} J & \mbox{if $ \bd{R}_i - \bd{R}_j = \pm ( \bd{\delta}_1 + \bd{\delta}_2 )$}
 \\
 J^{\prime}  & \mbox{if $ \bd{R}_i - \bd{R}_j = \pm  \bd{\delta}_1$ or $ \pm \bd{\delta}_2 $}
 \end{array},
 \right.
 \end{equation}
where 
 \begin{subequations}
 \begin{align}
\bm{\delta}_1 & =  \frac{b}{2} \bm{e}_x + \frac{c}{2} \bm{e}_y,
 \\
\bm{\delta}_2 & =  \frac{b}{2} \bm{e}_x - \frac{ c}{2} \bm{e}_y,
\end{align}
 \end{subequations}
so that  $\bd{\delta}_1 + \bd{\delta}_2  = b \bm{e}_x$ (see Fig. \ref{fig:lattice}). 
The DM vectors
$\bm{D}_{ij}$ are assumed to point in the $z$-direction and connect sites in the
directions $\pm \bm{\delta}_1$ and $\pm \bm{\delta}_2$,
 \begin{equation}
 \bm{D}_{ ij}  = D_{ij}  {\bm{e}}_z =  D ( \bd{R}_i - \bd{R}_j )  {\bm{e}}_z \; ,
 \end{equation}
with $D_{ij}  = - D_{ji}     $ given by
 \begin{equation}
 D_{ij} =  D ( \bd{R}_i - \bd{R}_j )   = \pm  D  \; 
\mbox{   if $ \bd{R}_i - \bd{R}_j = \pm  \bd{\delta}_1$ or $ \pm \bd{\delta}_2 $}.
 \end{equation}
For Cs$_2$CuCl$_4$, experimental estimates for $J$, $J^{\prime}$, $D$, and the
interlayer exchange coupling $J^{\prime \prime}$ are summarized in
Tab.~\ref{tab:cscucl}.
\begin{table}[tb]
 
\centering
\begin{ruledtabular}
\begin{tabular}{cc}
  Parameter & Value (meV)\\
\hline
 $J$&0.374(5)\\
 $J'$&0.128(5)\\
 $J''$&0.017(2)\\
 $D$&0.020(2)\\
\end{tabular}
\end{ruledtabular}
\caption{\label{tab:cscucl}
Accepted values~\cite{Coldea02, Veillette05a} 
of the in-plane exchange interactions $J$ and $J^{\prime}$,
the inter-plane interaction $J^{\prime \prime}$, and the
Dzyaloshinskii--Moriya interaction $D$ 
in Cs$_2$CuCl$_4$.  All comparisons between 
theory and experiment presented in this paper are based on these values.
Note that the largest exchange coupling
$J= 0.374~\mathrm{meV}$ corresponds to a temperature of $4.34 $ Kelvin.
}
\end{table}
Throughout this paper, we assume that the external magnetic field is perpendicular
to the plane of the lattice, $\bd{h} = h  {\bm{e}}_z$.   In the experimentally relevant regime
 $J^{\prime} < 2 J$ the spin system has then a unique classical ground state, as discussed
in the following section. If the magnetic field lies in the plane of the
lattice the ground state is more complicated and cannot be described within the framework
of the spin-wave expansion used in this paper.~\cite{Starykh10}

\subsection{Spin-wave expansion}
\label{sec:spinwave}

\subsubsection{General procedure}

\begin{figure}[tb]
 \centering
  \includegraphics[width=0.4\textwidth]{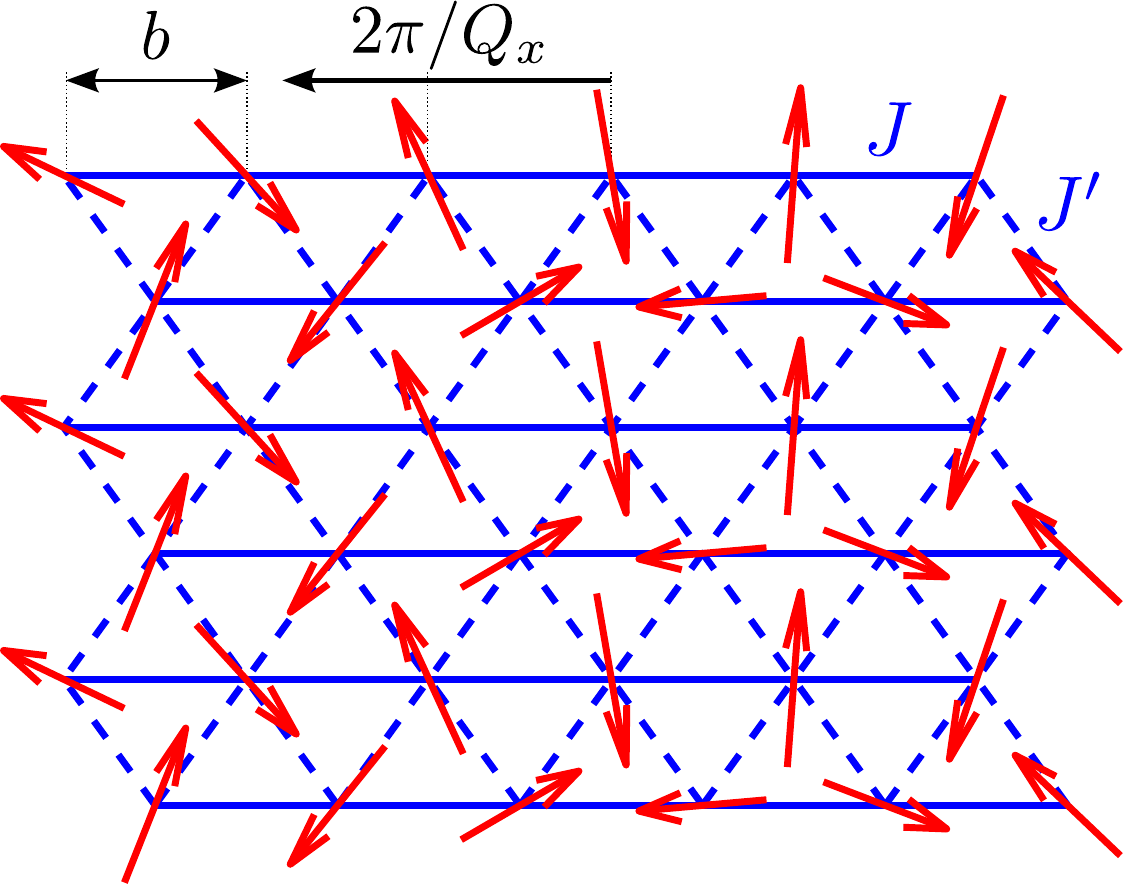}
 \caption{(Color online) Projection of the graph of the ``cone state'', which is the
classical ground state of the anisotropic triangular lattice antiferromagnet
for $J^{\prime} < 2 J$ and magnetic field perpendicular to the lattice plane. 
Note that the wavelength $2 \pi / Q_x$ of the spiral has the indicated length and
points along the $x$-axis in the direction of the arrow.}
 \label{fig:cone}
 \end{figure}

To set up the spin-wave expansion,
we should first identify the ground-state spin configuration
in the classical limit, where
the spin operators are replaced by three-component classical
vectors of length $S$, i.e., $\bd{S}_i \rightarrow S \hat{\bd{m}}_i$,
where $\hat{\bd{m}}_i$ are unit vectors which point in the direction
of the local magnetization.~\cite{Veillette05} 
In the experimentally relevant regime
$J^{\prime} < 2 J$ the classical ground  state of our model is a spiral
in the plane of the lattice, which is tilted towards the direction of the
magnetic field. In this so-called ``cone state'' the  magnetization 
points locally in the direction~\cite{Veillette05} 
\begin{equation}
 \hat{\bd{m}}_i  = 
 c_\theta   [ \cos ( \bd{Q} \cdot \bd{R}_i )    \bm{e}_x +  
   \sin ( \bd{Q} \cdot \bd{R}_i )  \bm{e}_y ] +
 s_\theta  \; \bm{e}_z ,
 \label{eq:spiral}
 \end{equation}
where we have introduced the abbreviation
 \begin{align}
 &s_{\theta } = \sin \theta \;,
&c_{\theta} = \cos \theta.
 \end{align}
This state is characterized by the opening angle $\theta$
of the cone and the wavevector $\bd{Q}$ of the spiral, as shown
in Fig.~\ref{fig:cone}.
To bosonize our spin Hamiltonian 
using the Holstein--Primakoff transformation,
we follow the general procedure outlined in Refs.~[\onlinecite{Schuetz03,Spremo05}] and
complement
the unit vectors $ \hat{\bd{m}}_i$ by
two orthogonal unit vector
$\bd{e}_i^{(1)}$ and $\bd{e}_i^{(2)}$ such that
$\{ \bd{e}_i^{(1)}, \bd{e}_i^{(2)}, \hat{\bd{m}}_i \}$ form a right-handed
triad of orthogonal unit vectors
for each site $\bd{R}_i$.
The transverse basis vectors 
$\bd{e}_i^{(1)}$ and $\bd{e}_i^{(2)}$
are defined only up to a local
$U(1)$ gauge transformation.~\cite{Schuetz03}
A convenient choice is
 \begin{align}
 \bd{e}^{(1)}_i
 & =    \sin ( \bd{Q} \cdot \bd{R}_i )    \bm{e}_x   
   - \cos ( \bd{Q} \cdot \bd{R}_i )  \bm{e}_y,
 \label{eq:ei1}
 \\
  \bd{e}^{(2)}_i & = 
s_\theta   [ \cos ( \bd{Q} \cdot \bd{R}_i )    \bm{e}_x +  
   \sin ( \bd{Q} \cdot \bd{R}_i )  \bm{e}_y ] -
 c_\theta  \; \bm{e}_z .
 \label{eq:ei2}
 \end{align}
Next we introduce the corresponding spherical basis vectors
 \begin{equation}
{\bd{e}}^{p}_i =
  {\bd{e}}^{(1)}_i + i p {\bd{e}}^{(2)}_i, \; \; \; p = \pm,
 \label{eq:eip}
\end{equation}
and expand the spin operators in this basis,
 \begin{equation}
 {\bd{S}}_i =
  S^{\parallel}_i \hat{\bd{m}}_i + {\bd{S}}^{\bot}_i,
 \end{equation}
with the transverse part given by
  \begin{equation}
  {\bd{S}}^{\bot}_i =\frac{1}{2} \sum_{ p = \pm } S_i^{-p}
  {\bd{e}}^{p}_i.
 \end{equation} 
The spin components are then bosonized using the Holstein--Primakoff transformation,
 \begin{subequations}
 \begin{align}
 S_i^{\parallel} & =  S- n_i,
 \\
 S_i^+ &=  \sqrt{2S} \sqrt{ 1 - \frac{ n_i}{2S}} b_i ,
 \\
 S_i^- &=  \sqrt{2S} b^{\dagger}_i \sqrt{ 1 - \frac{ n_i}{2S}}  ,
\end{align}
\end{subequations}
where $n_i = b^{\dagger}_i b_i$ and $ b_i$ are canonical boson operators.
The bosonized spin part of our Hamiltonian can then be written as
 \begin{equation}
 {H}_{\mathrm{spin}} = H_0 + H_{2 \parallel} + H_{4 \parallel } + {H}_{\bot} + 
{H}_{\parallel \bot},
 \label{eq:Hsdef}
 \end{equation}
where
 \begin{align}
  H_0 & = 
  \frac{S^2}{2} \sum_{  ij}   J^{\parallel} _{ij}  
  -  S \sum_{i  }  {\bd{h}}  \cdot \hat{\bd{m}}_i ,
  \label{eq:H0} 
 \\
 H_{2 \parallel} & = 
  - \frac{S}{2} \sum_{  ij}   J_{ij}^{\parallel}  
   ( n_i + n_j ) +   \sum_{i  }  {\bd{h}}  \cdot \hat{\bd{m}}_i n_i ,
  \label{eq:H2parallel} 
 \\
 H_{4 \parallel} & = 
   \frac{1}{2} \sum_{  ij}  J_{ij}^{\parallel}     n_i  n_j ,
  \label{eq:H4parallel} 
 \\
 {H}_{\bot}   & =  \frac{1}{2} \sum_{ i j  } 
 \left[ 
J_{ij} {\bd{S}}^{\bot}_i \cdot {\bd{S}}^{\bot}_j +
 \bd{D}_{ ij} \cdot (  {\bd{S}}^{\bot}_i \times {\bd{S}}^{\bot}_j ) \right]
 \nonumber
 \\
 & =  \frac{1}{8} \sum_{ ij} \sum_{ p p^{\prime}}
   J_{ij}^{ p p^{\prime}}
 S^{-p}_i S^{-p^{\prime}}_j,
 \label{eq:Hbot}
 \\
 {H}_{\parallel \bot} & = 
   \sum_{ i j } \bd{S}_i^{\bot} \cdot \left[
    (J_{ij}  - \bd{D}_{ij} \times ) \hat{\bd{m}}_j (S -n_j )  
- \delta_{ij} \bd{h}  
\right]
 \nonumber
 \\
 & = - 
   \sum_{ i j } \bd{S}_i^{\bot} \cdot \left[
    (J_{ij}  - \bd{D}_{ij} \times ) \hat{\bd{m}}_j n_j   
\right],
\label{eq:Hprime}
\end{align}
where in the last line we have used the fact that
the spin configuration in the classical ground state
satisfies~\cite{Schuetz03}
\begin{equation}
  \hat{\bd{m}}_i
  \times
  \Big[
  {\bd{h}} - S \sum_j ( J_{ij} -   \bd{D}_{ij} \times )  \hat{\bd{m}}_j
  \Big] = 0.
  \label{eq:classical}
\end{equation}
For simplicity we have introduced the following effective couplings
between longitudinal and transverse spin fluctuations,
 \begin{subequations}
 \begin{align}
 J_{ij}^{\parallel} & =  J_{ij}  
  \hat{\bd{m}}_i \cdot \hat{\bd{m}}_j +
 {\bd{D}}_{ij} \cdot (  \hat{\bd{m}}_i \times \hat{\bd{m}}_j ) ,
 \label{eq:Kparallel} 
\\
  J_{ij}^{ p p^{\prime}} & =  J_{ij} ( \bd{e}^p_i \cdot \bd{e}_j^{p^{\prime}} ) +
 {\bd{D}}_{ ij} \cdot   ( \bd{e}^p_i \times \bd{e}_j^{p^{\prime}} ).
 \label{eq:Kpp} 
\end{align}
  \end{subequations}
Defining $\varphi_i =  \bd{Q} \cdot \bd{R}_i$ and 
$\varphi_{ij} = \varphi_i - \varphi_j =  \bd{Q} \cdot ( \bd{R}_i - \bd{R}_j )$,
we obtain with our choice of basis
 \begin{subequations}
 \begin{align}
 J_{ij}^{\parallel} & =   J_{ ij}  [ s_{\theta}^2 + c_{\theta}^2 \cos  \varphi_{ij } ]
-   D_{ ij}  c_{\theta}^2 \sin \varphi_{ij}     ,
\label{eq:Jparallel}
 \\
 J_{ij}^{+-} & =   ( J_{ij}^{-+} )^{\ast} 
 \nonumber
 \\
& =  J_{ij}
 [  c_{\theta}^2 + ( 1 + s_{\theta}^2 ) \cos  \varphi_{ij } - 2 i s_{\theta }
 \sin \varphi_{ij } ]
 \nonumber
 \\
 & -    D_{ij}  [  ( 1 + s_{\theta}^2 ) \sin  \varphi_{ij } + 2 i s_{\theta }
 \cos \varphi_{ij } ]   ,
 \label{eq:Jpm}
 \\
 J_{ij}^{++} & =  J_{ij}^{--}  
 \nonumber
\\
&  =   -  c_{\theta}^2 \left[  J_{ij}    ( 1 -  \cos  \varphi_{ij } ) 
 +  D_{ij}      \sin  \varphi_{ij }  \right] .
 \label{eq:Jpp}
 \end{align}
\end{subequations}
It is convenient to introduce the Fourier transforms of the exchange couplings and the
DM interaction:
\begin{align}
 J_{\bd{k}} & =  \sum_{ \bd{R}} e^{ - i \bd{k} \cdot \bd{R}} J ( \bd{R} )
 \nonumber
 \\
 & =  2 J \cos ( k_x b ) + 4 J^{\prime} \cos ( {k_x b}/{2} ) 
\cos ({  k_y c}/{2} ) , 
 \\
 {{D}}_{\bd{k}} & =  \sum_{ \bd{R}} e^{ - i \bd{k} \cdot \bd{R}} {{D}} ( \bd{R} ) 
 \nonumber
 \\
 & =  - 4 i D  \sin ( {k_x b}/{2} ) 
\cos ( {  k_y c}/{2} ).
 \end{align}
We also define the combination
 \begin{equation}
 J_{\bd{k}}^D = J_{\bd{k}} - i D_{\bd{k}},
 \label{eq:FQdef}
 \end{equation}
which is real because the
Fourier transform of the
DM coupling $D_{\bd{k}}$ is purely imaginary.
The Fourier transforms of the effective couplings (\ref{eq:Jparallel}--\ref{eq:Jpp}) can then 
be written as
 \begin{subequations}
\begin{align}
 J^{\parallel}_{\bd{k}} & = 
s_{\theta}^2 J_{\bd{k}} + c_{\theta}^2
 \frac{ J^{D}_{ \bd{Q} + \bd{k}} + J^{D}_{\bd{Q} - \bd{k}} }{2} ,
 \label{eq:JparallelFT}
 \\
J^{+-}_{\bd{k}} = J^{-+}_{- \bd{k}}   & = 
c_{\theta}^2 J_{\bd{k}} + (1+ s_{\theta}^2)
 \frac{ J^{D}_{ \bd{Q} + \bd{k}} + J^{D}_{\bd{Q} - \bd{k}} }{2} 
 \nonumber
 \\
 & 
+ s_{\theta} \left[ J^{D}_{ \bd{Q} + \bd{k}} - J^{D}_{\bd{Q} - \bd{k}} \right],
 \\
J^{++}_{\bd{k}} = J^{--}_{\bd{k}}   & =  - c_{\theta}^2 \left[
J_{\bd{k}} -
 \frac{ J^{D}_{ \bd{Q} + \bd{k}} + J^{D}_{\bd{Q} - \bd{k}} }{2} \right] .
 \label{eq:JtransverseFT}
 \end{align}
\end{subequations}

\subsubsection{Classical ground state: Cone state}
\label{sec:classical}
To fix the parameters $\theta$ and $\bd{Q}$ which characterize the classical
ground state,
we 
substitute  Eq.~(\ref{eq:Jparallel}) into 
our expression (\ref{eq:H0}) for the classical ground-state energy and minimize
with respect to $\theta$ and $\bd{Q}$.
The classical ground-state energy can then be written as
 \begin{align}
 H_0 & =   N \frac{ S^2}{2} J^{\parallel}_{ \bd{k} =0} -  N S h s_\theta 
 \nonumber
 \\
 & =  N \frac{ S^2}{2} \left[   s_{\theta}^2   J_{\bd{k} =0} +  c_{\theta}^2    J^{D}_{\bd{Q}}  
 \right]  -  N S h s_\theta .
 \label{eq:Hclspiral}
\end{align}
Minimizing this with respect to $\theta$ we obtain
\begin{align}
  s_{\theta} \equiv \sin \theta = h/ h_c,
 \label{eq:tiltclassical}
 \end{align}
 where the critical magnetic field is given by
\begin{equation}
 h_c = S ( J^{D}_0 - J^{D}_{\bd{Q}}) = S( J_0 - J_{\bd{Q}} + i D_{\bd{Q}}).
 \label{eq:hcdef} 
\end{equation}
The wavevector of the spiral is 
obtained by minimizing $H_0$ in Eq.~(\ref{eq:Hclspiral})
with respect to $\bd{Q}$, which amounts to finding the minimum of the
real function $J^{D}_{\bd{Q}}  $. The wavevector of the spiral is
thus determined:
\begin{equation}
 \nabla_{\bd{Q}} J^D_{\bd{Q}} \equiv
\nabla_{\bd{k}} \left( J_{\bd{k}} - i D_{\bd{k}} \right)_{ \bd{k} = \bd{Q} } =0.
 \label{eq:Qcond} 
\end{equation}
Anticipating that this condition leads to 
a spiral along  the $x$-axis, $\bd{Q} = Q_x \bd{e}_x$, and using
the above expressions for
$J_{\bd{k}}$ and $D_{\bd{k}}$, it is easy to show that
Eq.~(\ref{eq:Qcond}) reduces to the following 
transcendental equation for $x = Q_x b$:
 \begin{equation}
 \cos \left( \frac{x}{2} \right) = - \frac{J^{\prime}}{2J} -
 \frac{D}{2 J} \cot \left( \frac{x}{2} \right).
 \end{equation}
Note that for $D=0$ and isotropic exchange couplings ($J^{\prime} = J$) 
this condition reduces to $\cos ( x/2 ) = - 1/2$, implying $x = Q_x b = 4 \pi /3$, 
which describes
the usual $120^{\circ}$ ground state of an isotropic triangular lattice antiferromagnet.

\subsubsection{Magnon dispersion}

\begin{figure}[tb]
\centering
  \includegraphics[width=0.45\textwidth]{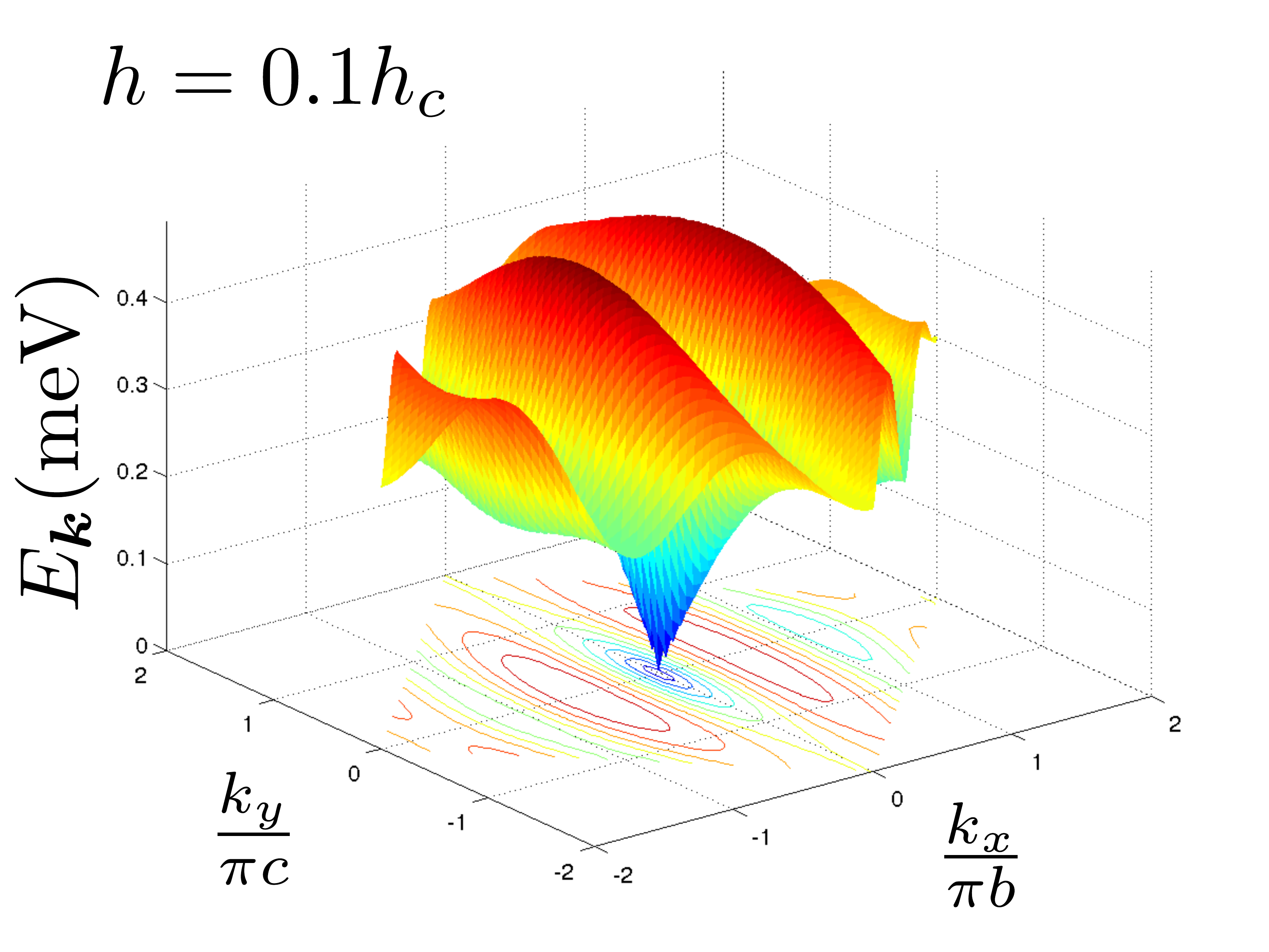}
 \caption{(Color online)
Graph of the magnon dispersion $E_{\bd{k}}$ of the 
anisotropic triangular lattice antiferromagnet Cs$_2$CuCl$_4$ with
$J^{\prime} /  J = 0.34$, Dzyaloshinskii--Moriya anisotropy $D/J= 0.054$, and
for a magnetic field $h=0.1 h_c$;
see Eq.~(\ref{eq:magdispersion}).
Note that the  $U(1)$-symmetry of the Hamiltonian guarantees that for $\bd{k} =0$
the magnon dispersion is gapless.\cite{Veillette05}
}
 \label{fig:magdisp}
 \end{figure}

\begin{figure}[tb]
\centering
    \includegraphics[width=0.40\textwidth]{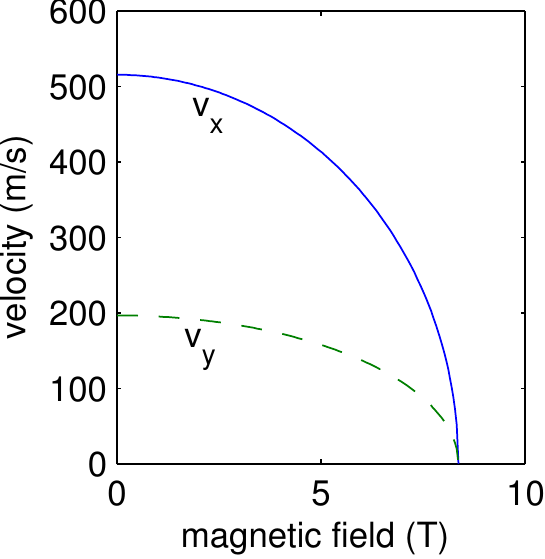}
 \caption{(Color online)
Plot of the linear spin-wave result for the magnon velocities of  Cs$_2$CuCl$_4$
in the two principal directions of the lattice as a function of the magnetic field.
The relevant values of $J$, $J^{\prime}$, and $D$ are given in
Tab.~\ref{tab:cscucl}.
It turns out that for an arbitrary magnetic field  the magnon velocities  
are small compared with the  velocities $c_{\lambda}$ of acoustic phonons
in Cs$_2$CuCl$_4$.
}
 \label{fig:magdispa}
 \end{figure}

To calculate the magnon spectrum to leading order in the $1/S$ expansion,
we may approximate
$S^{+}_i \approx \sqrt{2S} b_i$ and
$S^{-}_i \approx \sqrt{2S} b^{\dagger}_i$, so that the transverse part of
our spin-wave Hamiltonian is approximated by $H_{\bot} \approx H_{2 \bot}$ with
 \begin{align}
 H_{2 \bot} & =  \frac{S}{4} \sum_{ij}
 \bigl[ J_{ij}^{+-} b^{\dagger}_i b_j  + J_{ij}^{-+}  b_i  b_j^{\dagger}
 \nonumber\\&  \hspace{8mm} 
+  J_{ ij}^{++}  b^{\dagger}_i b^{\dagger}_j +
J_{ ij}^{--} b_i b_j 
 \bigr].
 \label{eq:H2bot}
 \end{align}
The magnon spectrum can then be obtained by diagonalizing the
quadratic boson Hamiltonian
 $H_2 =  H_{2 \parallel} + H_{2 \bot}$.
Combining Eqs.~(\ref{eq:H2parallel}) and (\ref{eq:H2bot}) we  obtain
 \begin{align}
H_2 & =    \frac{S}{2} \sum_{  ij}  \Bigl\{ 
 -  J_{ij}^{\parallel} ( n_i + n_j ) +
\frac{1}{2} \bigl[ J_{ij}^{+-} b^{\dagger}_i b_j  + J_{ij}^{-+}  b_i  b_j^{\dagger}
 \nonumber
 \\
 &  \hspace{3mm} +  J_{ ij}^{++}  b^{\dagger}_i b^{\dagger}_j 
 + J_{ ij}^{--} b_i b_j 
 \bigr] \Bigr\}
+   h s_{\theta} \sum_{i  }  n_i .
 \hspace{7mm}
 \label{eq:H2SW}
\end{align}
Introducing the Fourier transform of the boson operators via
 \begin{equation}
 b_i = \frac{1}{\sqrt{N}} \sum_{\bd{k}} e^{ i \bd{k} \cdot {\bd{R}}_i } b_{\bd{k}},
 \end{equation}
we obtain in momentum space,
 \begin{align}
 H_2 & =  \sum_{\bd{k}}  \Bigl\{  A_{\bd{k}}
b^{\dagger}_{\bd{k}} b_{\bd{k}} 
 + \frac{ B_{\bd{k}}}{2} \bigl[ 
 b^{\dagger}_{\bd{k}} b^{\dagger}_{-\bd{k}} +
b_{-\bd{k}} b_{\bd{k}} \bigr] \Bigr\},
 \label{eq:H2bb}
 \end{align} 
where $A_{\bd{k}} = A_{\bd{k}}^{+} + A_{\bd{k}}^{-}$, with
 \begin{subequations}
 \begin{align}
 A^{+}_{\bd{k}} & =    - B_{\bd{k}}
- S
 \left[  J^{D}_{\bd{Q}} - \frac{ J^{D}_{ \bd{Q} + \bd{k}} +   J^{D}_{ \bd{Q} - \bd{k}} }{2} 
\right], 
 \label{eq:Akdef}
 \\
 A^-_{\bd{k}} & =      S s_{\theta}
\frac{  J^{D}_{ \bd{Q} + \bd{k}} -   J^{D}_{ \bd{Q} - \bd{k}} }{2},
 \label{eq:Akminusdef}
\\
  B_{\bd{k}} & =      - 
\frac{S}{2}     c_{\theta}^2
 \left[  J_{\bd{k}} - \frac{ J^{D}_{ \bd{Q} + \bd{k}} +   J^{D}_{ \bd{Q} - \bd{k}} }{2} \right].
 \label{eq:Bkdef}
\end{align}
 \end{subequations}
These coefficients are real and have the symmetries
$A^{\pm}_{-\bd{k}} = \pm A^\pm_{  \bd{k}}$ and
 $B_{\bd{k}} = B_{ - \bd{k}}$.
Note that
 \begin{subequations}
 \begin{align}
 B_0  & =  - \frac{S}{2} c_{\theta}^2 [ J_0 - J^{D}_{\bd{Q}} ] = - \frac{c_{\theta}^2}{2} h_c,
 \label{eq:B0}
 \\
 A_0 & =  \frac{c_{\theta}^2}{2} h_c,
 \label{eq:A0}
 \end{align}
 \end{subequations}
which implies that the magnon dispersion is gapless at $\bd{k}=0$, as
required by the $U(1)$-symmetry of the Hamiltonian.
Note also that for small $\bd{k}$ the linear coefficient in the Taylor expansion of
the 
 antisymmetric coefficient
$A^{-}_{\bd{k}}$  vanishes due to Eqs.~(\ref{eq:FQdef}) and (\ref{eq:Qcond}), so that
 \begin{equation}
 A^{-}_{\bd{k}}
=  {\mathcal{O}} ( \bd{k}^3 ) .
 \label{eq:Aminussmall}
 \end{equation}
To obtain the magnon dispersion, we diagonalize the 
Hamiltonian (\ref{eq:H2bb}) using the following canonical (Bogoliubov) transformation,
\begin{align}
 \left( \begin{array}{c}
  b_{\bd{k}} \\
 b^{\dagger}_{-\bd{k}}
 \end{array}
 \right) = \left( \begin{array}{cc} u_{\bd{k}} & - v_{\bd{k}} \\
 -  v_{\bd{k}} & u_{\bd{k}}  \end{array} \right) 
\left( \begin{array}{c}
  \beta_{\bd{k}} \\
 \beta^{\dagger}_{-\bd{k}}
 \end{array}
 \right),
 \label{eq:Bogoliubov}
\end{align}
where
 \begin{equation}
 u_{\bd{k}} = \sqrt{ \frac{ A^+_{\bd{k}} + \epsilon_{\bd{k}} }{ 2 \epsilon_{\bd{k}} } } 
 \; \;   , \; \;
 v_{\bd{k}} = \frac{ B_{\bd{k}}}{ | B_{\bd{k}}    | }
\sqrt{ \frac{ A^+_{\bd{k}} - \epsilon_{\bd{k}} }{ 2 \epsilon_{\bd{k}} } } ,
 \end{equation}
and
 \begin{equation}
 \epsilon_{\bd{k}} = \sqrt{ (A^+_{\bd{k}})^2 -  B_{\bd{k}}^2 }.
 \end{equation}
The energy $\epsilon_{\bd{k}} =  \epsilon_{ - \bd{k}}$ 
is even under $\bd{k} \rightarrow - \bd{k}$;
the full magnon dispersion is
 \begin{equation}
 E_{\bd{k}} =  
\epsilon_{\bd{k}} + A^-_{\bd{k}} = \sqrt{ (A^+_{\bd{k}})^2 -  B_{\bd{k}}^2 } + A^-_{\bd{k}},
 \label{eq:magdispersion}
 \end{equation}
which does not have any definite
symmetry with respect to $\bd{k} \rightarrow - \bd{k}$.
The diagonalized form of the magnon Hamiltonian reads
 \begin{equation}
 H_2 = \sum_{\bd{k}} \left[ E_{\bd{k}} \beta^{\dagger}_{\bd{k}} \beta_{\bd{k}} +
 \frac{ \epsilon_{\bd{k}} - A^+_{\bd{k}}}{2} \right].
 \end{equation}
A graph of the magnon dispersion $E_{\bd{k}}$ is shown in
Fig.~\ref{fig:magdisp}.
Using the fact that according to Eq.~(\ref{eq:Aminussmall})
the term $A^{-}_{\bd{k}}$ can be neglected for small $\bd{k}$, the leading
long-wavelength limit of the magnon dispersion is
  \begin{equation}
 E_{\bd{k}} =   \sqrt{v_x^2 k_x^2 + v_y^2 k_y^2} + {\mathcal{O}} ( \bd{k}^3 ) = v ( \hat{\bd{k}} ) | \bd{k} |  + {\mathcal{O}} ( \bd{k}^3 ) ,
 \label{eq:Eksmall}
 \end{equation}
with direction-dependent magnon velocity
\begin{equation}
v ( \hat{\bd{k}} ) = \sqrt{ v_x^2 \hat{k}_x^2 + v_y^2 \hat{k}_y^2 },
 \label{eq:vkhat}
 \end{equation}
where $\hat{\bd{k}} = \bd{k} / | \bd{k} |$.   
 In Fig.~\ref{fig:magdispa} we show a
graph of the linear spin-wave result of the 
two principal magnon velocities $v_x$ and $v_y$ as a function of
the external magnetic field for Cs$_2$CuCl$_4$, where
 $J, J^{\prime}$, and $D$   have the values given in 
Tab.~\ref{tab:cscucl}.
\subsubsection{Interactions between magnons}

The terms in the transverse Hamiltonian $H_{\bot}$ involving more than
two boson operators and the contributions
$H_{4 \parallel}$ and $H_{\parallel \bot}$ to the magnon Hamiltonian
given in Eqs.~(\ref{eq:H4parallel}) and (\ref{eq:Hprime}) describe various 
interaction processes between magnons.
As compared with the $S$-dependence of the free magnon dispersion
$E_{\bd{k}} \propto S$,
the interaction contains higher orders
of the small parameter $1/S$, so that at first sight it seems that
for large $S$ we can simply neglect interaction effects.
It turns out, however, that for the proper calculation
of  the long-wavelength limit of the ultrasonic attenuation rate
the  leading contribution $H_3$ to the mixing term $H_{\parallel \bot}$
involving three magnon operators cannot be neglected.
This part of the magnon Hamiltonian can be written as
 \begin{equation}
 H_3 = - c_{\theta}  \frac{ \sqrt{2S}}{2 i }
 \sum_{i j}  \left[ K_{ij} b^{\dagger}_i b^{\dagger}_j b_j - K_{ij}^{\ast} b^{\dagger}_j  b_j b_i  
\right],
 \label{eq:H3}
 \end{equation}
where we have defined
 \begin{align}
  K_ {ij} & =  s_{\theta} \left[ J_{ij} ( 1 - \cos \varphi_{ij} ) + D_{ij} \sin \varphi_{ij} \right]
 \nonumber
 \\
 &   + i \left[ J_{ij} \sin \varphi_{ij} + D_{ij} \cos \varphi_{ij} \right].
 \label{eq:Kijdef}
 \end{align} 
In momentum space $H_3$ can be written as
 \begin{align}
H_3  & =  \frac{1}{\sqrt{N} }
 \sum_{ \bd{k}_1 \bd{k}_2 \bd{k}_3} \delta_{ \bd{k}_1 +
 \bd{k}_2 + \bd{k}_3  , 0 } \Bigl[
 \nonumber
 \\
 &  \frac{1}{ 2!}
\Gamma_3^{ {b}^{\dagger} {b}^{\dagger} b }
 ( \bd{k}_1 , \bd{k}_2 ; \bd{k}_3 )
 b^{\dagger}_{  - \bd{k}_1 } b^{\dagger}_{  - \bd{k}_2 }
 b_{  \bd{k}_3 }  
 \nonumber
 \\
 & +  \frac{1}{ 2!} 
  \Gamma_3^{ {b}^{\dagger} bb }
 ( \bd{k}_1 ; \bd{k}_2 , \bd{k}_3 )
 b^{\dagger}_{ - \bd{k}_1 } b_{  \bd{k}_2 }
 b_{ \bd{k}_3 }
\Bigr],
 \label{eq:H3hp}
 \end{align}
with  the vertices given by
 \begin{subequations}
 \begin{align}
\Gamma_3^{ {b}^{\dagger} {b}^{\dagger} b }
 ( \bd{k}_1 , \bd{k}_2 ; \bd{k}_3 ) & = 
   - c_{\theta}  \frac{ \sqrt{2S}}{2 i }
 \left[ K_{-\bd{k}_1} + K_{ - \bd{k}_2 } \right], 
 \label{eq:Gammabbb1}
\\
 \Gamma_3^{ {b}^{\dagger} b b }
 ( \bd{k}_1 ; \bd{k}_2 , \bd{k}_3 ) & = 
    c_{\theta}  \frac{ \sqrt{2S}}{2 i }
 \left[ K_{\bd{k}_2} + K_{  \bd{k}_3 } \right],
 \label{eq:Gammabbb2}
\end{align}
 \end{subequations}
where $K_{\bd{k}}$ is the Fourier transform of the
function $K_{ij}$ defined in Eq.~(\ref{eq:Kijdef}):
 \begin{equation}
 K_{\bd{k}} = s_{\theta} 
\left[  J_{\bd{k}} - \frac{ J^{D}_{ \bd{Q} + \bd{k}} +   J^{D}_{ \bd{Q} - \bd{k}} }{2} \right] -
\frac{  J^{D}_{ \bd{Q} + \bd{k}} -   J^{D}_{ \bd{Q} - \bd{k}} }{2}.
 \label{eq:Kkdef}
 \end{equation}
For later reference we note that
 \begin{equation}
 K_0 = s_{\theta} ( J_0 - J_{\bd{Q}}^D) = s_{\theta} \frac{h_c}{S}.
 \label{eq:K0lim}
 \end{equation}

\subsection{Magnon-phonon interactions}
\label{sec:magpho}

\subsubsection{Modeling the derivatives  of the exchange couplings
in Cs$_2$CuCl$_4$
}

Substituting the gradient expansions (\ref{eq:Jgrad}) and (\ref{eq:Dgrad})
for the exchange and DM couplings
into the spin-phonon Hamiltonian $H^{\mathrm{pho}}_{\mathrm{spin}}$
defined in Eq.~(\ref{eq:Hs}), we obtain an expansion
 of $H^{\mathrm{pho}}_{\mathrm{spin}}$
in powers of the phonon operators of the form (\ref{eq:Hsp}).
Since the exchange interactions are large compared with the
DM interactions, we shall neglect the
dependence of the DM couplings $\bd{D}_{ij}$
in the phonon coordinates. The $n$-phonon part of our spin-phonon Hamiltonian can be written as
\begin{align}
 H^{n \mathrm{pho}}_{\mathrm{spin}}  & =   \frac{1}{2} \sum_{ij} U^{(n)}_{ij} {\bd{S}}_i \cdot {\bd{S}}_j,
 \label{eq:Usp} 
\end{align}
where the coupling functions involving $n=1$ and $n=2$ phonons are
 \begin{align}
U^{(1)}_{ ij} & =    ({\bd{X}}_{ij} \cdot \nabla_{\bd{r}}  ) \left. J ( \bd{r} )
\right|_{ \bd{r} = \bd{R}_{ij} } \equiv {\bd{X}}_{ij} \cdot \bd{J}^{(1)}_{ij},
 \label{eq:U1def}
\\
 U^{(2)}_{ ij} & =  \frac{1}{2}  ({\bd{X}}_{ij} \cdot \nabla_{\bd{r}}  )^2 \left. J ( \bd{r} )
\right|_{ \bd{r} = \bd{R}_{ij} } \equiv 
\frac{1}{2}  {\bd{X}}_{ij}^T  \mathbf{J}^{(2)}_{ij}  {\bd{X}}_{ij} , \hspace{9mm}
 \label{eq:U2def}
 \end{align}
with the vector   $\bd{J}^{(1)}_{ij}$ and the tensor 
$\mathbf{J}^{(2)}_{ij}$ defined by
\begin{align}
  \bd{J}^{(1)}_{ij}  & \equiv   
\bd{J}^{(1)} ( \bd{R}_{ij} ) 
=\mathbf{\nabla}_{\bd{r}} 
\left. J ( \bd{r} )
\right|_{ \bd{r} = \bd{R}_{ij} },
\\
{[} \mathbf{J}^{(2)}_{ ij } {]}_{\alpha \beta}
 & \equiv   
{[} \mathbf{J}^{(2)} ( \bd{R}_{ij} ) {]}_{\alpha \beta}
=
 \left.
\frac{ J ( \bd{r} ) }{ \partial r_{\alpha} 
 \partial r_{\beta} } \right|_{ \bd{r} = \bd{R}_{ij} }.
 \label{eq:J2derivdef}
 \end{align}
The Fourier transforms of functions (\ref{eq:U1def}) and(\ref{eq:U2def}) are
 \begin{align}
 U^{(1)}_{ \bd{k} , \bd{k}^{\prime}} & =  \frac{1}{N} \sum_{ ij}
 e^{ - i \bd{k} \cdot \bd{R}_i - i    \bd{k}^{\prime} \cdot \bd{R}_j  }  U^{(1)}_{ij}
  \nonumber
 \\
 & =  - \frac{1}{\sqrt{N}} \bd{X}_{ \bd{k} + \bd{k}^{\prime}} \cdot (
 \bd{J}^{(1)}_{\bd{k} } + \bd{J}^{(1)}_{ \bd{k}^{\prime} } ),
 \label{eq:U1FT}
 \\
 U^{(2)}_{\bd{k} , \bd{k}^{\prime}} & = 
 \frac{1}{N} \sum_{ ij}
 e^{ - i \bd{k} \cdot \bd{R}_i - i    \bd{k}^{\prime} \cdot \bd{R}_j  }  U^{(2)}_{ij}
  \nonumber
 \\
&   \hspace{-10mm} =
\frac{1}{2N} \sum_{\bd{q}} 
 \bd{X}_{  \bd{q}}^{\dagger}  ( \mathbf{J}^{(2)}_{ \bd{k}} - \mathbf{J}^{(2)}_{ \bd{k} + \bd{q}}
+ \mathbf{J}^{(2)}_{ \bd{k}^{\prime}} - \mathbf{J}^{(2)}_{ \bd{k}^{\prime} + \bd{q}} ) 
 \bd{X}_{ \bd{q} + \bd{k} + \bd{k}^{\prime} },
 \label{eq:U2mat}
 \end{align}
where 
\begin{align}
\bd{J}^{(1)}_{\bd{k}} & = \sum_{\bd{R}} e^{ - i \bd{k} \cdot \bd{R}} \bd{J}^{(1)} ( \bd{R} ),
 \label{eq:JnFT}
 \end{align}
and similarly for $\mathbf{J}^{(2)}_{ \bd{k}}$.
Note that the Fourier transform $\bd{J}^{(1)}_{\bd{k}}$ of the gradient has the symmetries
 \begin{equation}
\bd{J}^{(1)}_{ - \bd{k}} = - \bd{J}^{(1)}_{  \bd{k}} = (\bd{J}^{(1)}_{\bd{k}})^{\ast},
\label{eq:symmft}
 \end{equation}
which implies that $\bd{J}^{(1)}_{\bd{k}}$ is purely imaginary and vanishes for $\bd{k} =0$. In contrast, the second derivative tensor is an even function of $\bd{k}$,
\begin{equation}
\mathbf{J}^{(2)}_{ - \bd{k}} = \mathbf{J}^{(2)}_{ \bd{k}},
\label{eq:symmft2}
 \end{equation}
and thus has a finite limit at zero momentum.
In general, in-plane lattice deformations in
Cs$_2$CuCl$_4$ induce changes in both couplings $J$ and $J^{\prime}$
as indicated in  Fig.~\ref{fig:lattice}.
In principle one should use four independent parameters to
describe  the changes of  $J$ and $J^{\prime}$
in response to the displacement of the lattice sites
in two independent in-plane directions.
To reduce the parameter space, we note that
usually the exchange couplings originate from an electronic exchange path 
where orbitals of neighboring atoms overlap. 
For Cs$_2$CuCl$_4$ 
the exchange is mediated by the chlorine atoms 
located in between the copper atoms,
as shown in Fig.~\ref{fig:lattice}.  
In a rough approximation where only the coupling of a 
few atoms is considered, the exchange couplings
do not change to first order in the displacements 
when one of the copper atoms moves up or down. 
Therefore one can neglect the variation of $J$ with respect to displacements 
in the $y$-direction and assume that $J$ only
changes appreciably  if the atoms move in the $x$-direction. 
By a similar argument one can also neglect the dependence of the coupling $J'(\bd r)=J'(r,\varphi)$ on the polar angle $\varphi$, setting
\begin{equation}
 \frac{\partial J'}{\partial x} =\cos \varphi_0 \frac{ \partial J'}{\partial r } \; \; , \; \; 
\frac{ \partial J'}{\partial y} =\sin \varphi_0 \frac{ \partial J'}{\partial r }\;,
\label{eq:derivatives}
\end{equation}
where  $\sin\varphi_0=c/(2 d)$. Here $d= \sqrt{ b^2 + c^2}/2$ 
is the relevant bond length, and
$\partial J' / \partial r $ 
is the derivative of the coupling $J'$ with respect to the distance between the 
spins.  With these approximations we obtain the following expression
for the gradient $\bd{J}^{(1)}_{\bd{k}}$
of the exchange couplings in  Cs$_2$CuCl$_4$:
 \begin{align}
 \bd{J}^{(1)}_{\bd{k}} 
 & =  - 2 i   \frac{ \partial J}{\partial x } \bd{e}_x  \sin ( k_x b )
 \nonumber
 \\
 &   - 4 i \frac{\partial J^{\prime}}{\partial r } \Bigl[
   \bd{e}_x   \cos \varphi_0 \sin \left( {k_x b}/{2} \right)
 \cos \left( {k_y c}/{2} \right) 
 \nonumber
 \\
 &  \hspace{12mm}
 +  \bd{e}_y    \sin \varphi_0 \cos ( {k_x b}/{2} )
 \sin ( {k_y c}/{2} ) \Bigr]. 
 \label{eq:J1vec}
 \end{align}
Similarly, we obtain for the matrix elements of the second derivative
tensor $\mathbf{J}^{(2)}_{\bd{k}}={\mathbf{J}^{(2)}_{\bd{k}}}^T$,
\begin{subequations}
\begin{align}
 \bigl[\mathbf{J}_{\bd k}^{(2)}\bigr]_{11}&=2\cos(k_xb)\frac{\partial^2J}{\partial x^2}
\notag\\&
+4\cos^2\varphi_0 \cos({k_x b}/{2})\cos ( {k_y c}/{2} ) \frac{\partial^2J'}{\partial r^2}\;,
 \label{eq:J2tensa}\\
 \bigl[\mathbf{J}_{\bd k}^{(2)}\bigr]_{12}&=4\cos \varphi_0 \sin\varphi_0
\sin({k_x b}/{2})\sin ( {k_y c}/{2} )  \frac{\partial^2J'}{\partial r^2}\;,
 \label{eq:J2tensb}\\
 \bigl[\mathbf{J}_{\bd k}^{(2)}\bigr]_{22}&=4\sin^2\varphi_0\cos({k_x b}/{2})
\cos ( {k_y c}/{2} )\frac{\partial^2J'}{\partial r^2}\;.
 \label{eq:J2tensc}
\end{align}
\end{subequations}
Due to the $U(1)$-symmetry of the Hamiltonian, 
the vector $ \bd{J}^{(1)}_{\bd{k}}$ and  the tensor 
$\mathbf{J}^{(2)}_{\bd{k}}$ are only fixed up to a phase that depends on the choice of triad  $\{ \bd{e}_i^{(1)}, \bd{e}_i^{(2)}, \hat{\bd{m}}_i \}$; see Ref. [\onlinecite{Schuetz03}]
for a discussion of this point.
Note that the geometry 
shown in Fig. \ref{fig:lattice} suggests 
that the couplings will become stronger if the bond length is enlarged;
at first sight this counterintuitive behavior
originates from the fact that the bond angles between the chlorine atoms become
larger with  increasing bond length, 
which usually results in a 
stronger coupling.\footnote{K. Foyevtsova, private communication} 
Later we will see that only the squares of the derivatives enter in the final result 
for the velocity shifts and the ultrasonic attenuation rates, so that
one cannot directly determine
the sign of the change of the coupling strengths 
from the ultrasound measurements.

\subsubsection{Phonon renormalization in the classical spin background}

Expanding the spin operators in Eq.~(\ref{eq:Usp})
in powers of Holstein--Primakoff bosons, we obtain
a double expansion of the spin-phonon Hamiltonian in
powers of phonon operators and magnon operators.
For our purpose, it is sufficient to retain only terms up to two phonon
operators. After expressing the spin operators
in Eq.~(\ref{eq:Usp}) in terms of Holstein--Primakoff bosons,
to leading order for large $S$, we generate also 
pure phonon contributions of the form
 \begin{equation}
 H^{n \mathrm{pho}}_{0} = \frac{S^2}{2} \sum_{ij} U^{(n)}_{ij} \hat{\bd{m}}_i \cdot \hat{\bd{m}}_j,
 \end{equation}
which describe motion of $n$ phonons in the classical spin background.
In particular,
the one-phonon term $H^{\mathrm{1pho}}_{0}$  can be written as
 \begin{equation}
 H^{\mathrm{1pho}}_{0} = \frac{S^2}{2} \left[ s_{\theta}^2 U^{(1)}_{0,0} + \frac{c_{\theta}^2}{2} 
 \left( U^{(1)}_{ - \bd{Q} , \bd{Q}} + U^{(1)}_{ \bd{Q} , - \bd{Q}}  \right) \right].
\label{eq:H10}
 \end{equation}
But from Eqs.~(\ref{eq:U1FT}) and (\ref{eq:symmft}) we see that
$U^{(1)}_{0,0} = U^ {(1)}_{ - \bd{Q} , \bd{Q} } =0$, implying
$H^{\mathrm{1pho}}_{0} =0$. 
On the other hand, the corresponding  two-phonon part
 \begin{equation}
H^{\mathrm{2pho}}_{0} = \frac{S^2}{2} \left[ s_{\vt}^2 U^{(2)}_{0,0} + \frac{c_{\vt}^2}{2} 
 \left( U^{(2)}_{ - \bd{Q} , \bd{Q}} + U^{(2)}_{ \bd{Q} , - \bd{Q}}  \right) \right]
 \label{eq:H2phozero}
 \end{equation}
is finite and yields an important contribution to the magnetic 
field dependence of the elastic  constants.
Inserting the explicit expression for the
Fourier transform
$U^{(2)}_{\bd{k} , \bd{k}^{\prime}}$
of the second derivative couplings  
$U^{(2)}_{ij}$ [see Eqs.~(\ref{eq:U2def}) and (\ref{eq:U2mat})]
and using  the symmetry \Ref{eq:symmft2} we see that Eq.~(\ref{eq:H2phozero})
can be written as
\begin{equation}
H^{\mathrm{2pho}}_{0} =\frac{M}{2}\sum_{{\bd k}\lambda}
\Sigma_0^{\mathrm{pho}}  ( \bd k, \lambda ) X_{ - \bd{k} \lambda}
  X_{ \bd{k} \lambda}\;,\label{eq:H20}
\end{equation}
where the zeroth order contribution 
to the phonon self-energy
(in an expansion
in powers of the magnon operators) is given by
\begin{equation}
\Sigma_0^{\mathrm{pho}}  ( \bd k, \lambda ) 
=\frac{S^2}{M}{\bd e}_{{\bd k}\lambda}^{\dagger} \left[ s_\vt^2 \left({\mathbf J}_0^{(2)}-{\mathbf J}_{\bd k}^{(2)}\right)+c_\vt^2 \mathbf{J}^{(2+)}_{ \bd{Q}  , \bd{k}} \right]{\bd e}_{{\bd k}\lambda}\;.
 \label{eq:sigmapho0}
\end{equation}
Here we have  defined
 \begin{subequations}
 \begin{align}
\mathbf{J}^{(2+)}_{ \bd{k}  , \bd{Q}} & =  \mathbf{J}^{(2)}_{ \bd{k}}
 -
\frac{  \mathbf{J}^{(2)}_{ \bd{k}  + \bd{Q}}
+ \mathbf{J}^{(2)}_{ \bd{k}  - \bd{Q}}}{2}\;.\label{eq:J2p}
 \end{align}
For later convenience we also introduce the notation
\begin{align}
\mathbf{J}^{(2-)}_{ \bd{k}  , \bd{Q}} & = \frac{  \mathbf{J}^{(2)}_{ \bd{k}  + \bd{Q}}
- \mathbf{J}^{(2)}_{ \bd{k}  - \bd{Q}}}{2}.
\label{eq:J2m}
 \end{align}
\end{subequations}
Keeping in mind that $ \mathbf{J}^{(2)}_{ \bd{k}}$ is an even function of $\bd{k}$, 
we see that $\mathbf{J}^{(2+)}_{ \bd{k}  , \bd{Q}}$ is even while 
$\mathbf{J}^{(2-)}_{ \bd{k}  , \bd{Q}}$ is odd under
$\bd{k} \rightarrow - \bd{k}$,
\begin{equation}
 \mathbf{J}^{(2\pm)}_{ \bd{k}, \bd{Q} } =
\pm \mathbf{J}^{(2\pm)}_{ -\bd{k} , \bd{Q}}\;.\label{eq:symmJ2}
\end{equation}

\subsubsection{One-phonon one-magnon hybridization}
\label{subsec:oneone}

The part $H^{\mathrm{1pho}}_{\mathrm{1mag}}$ 
of our spin-phonon Hamiltonian involving
one phonon and one magnon operator is given by
 \begin{align}
 H^{\mathrm{1pho} }_{\mathrm{1mag}} & =  \frac{ (2S)^{3/2}}{4} \sum_{ij} U^{(1)}_{ ij}
 \left[ b_i^{\dagger} ( \bd{e}_i^+ \cdot \hat{\bd{m}}_j ) +
 b_i ( \bd{e}_i^- \cdot \hat{\bd{m}}_j )  \right].
 \end{align}
Using 
 \begin{equation}
\bd{e}_i^\pm \cdot \hat{\bd{m}}_j = c_{\theta}   \sin \varphi_{ij} \mp i 
 c_{\theta} s_{\theta}  \left[ 1 -  \cos \varphi_{ij} \right],
\end{equation}
and Fourier transforming to momentum space, we obtain
 \begin{equation}
 H^{\mathrm{1pho}}_{\mathrm{1mag}} = \sum_{\bd{k}} \bd{\Gamma}_{\bd{k}}^{Xb} \cdot \left( 
 \bd{X}_{ - \bd{k}} b_{\bd{k}} + \bd{X}_{\bd{k}} b^{\dagger}_{\bd{k}} \right),
 \label{eq:H1p1m} 
\end{equation}
with the hybridization vertex given by
 \begin{align}
 \bd{\Gamma}_{\bd{k}}^{Xb} & =   \frac{i}{4} ( 2 S )^{3/2} c_{\theta}   
 \bigl[    \bd{J}^{(1+)}_{ \bd{k}, \bd{Q}}
 +    s_{\theta}  \bd{J}^{(1-)}_{ \bd{k}, \bd{Q}} \bigr].
 \label{eq:Gamma1p1m}
 \end{align}
Here we have defined 
 \begin{subequations}
 \begin{align}
 \bd{J}^{(1+)}_{\bd{k} , \bd{Q}} & =      \bd{J}^{(1)}_{\bd{Q}} - 
\frac{ \bd{J}^{(1)}_{\bd{k} + \bd{Q}} - \bd{J}^{(1)}_{\bd{k} - \bd{Q}}}{2}  ,
 \label{eq:gsdef}
 \\
 \bd{J}^{(1-)}_{\bd{k} , \bd{Q}} & =  \bd{J}^{(1)}_{\bd{k}} - 
\frac{ \bd{J}^{(1)}_{\bd{k} + \bd{Q}} + \bd{J}^{(1)}_{\bd{k} - \bd{Q}}}{2} .
 \label{eq:gadef} 
\end{align}
 \end{subequations}
For fixed $\bd{Q}$ the functions
$\bd{J}_{\bd{k} , \bd{Q}}^{(1 \pm)}$
have the symmetries
 \begin{equation}
 \bd{J}_{ - \bd{k} , \bd{Q}}^{(1\pm)}  = \pm   \bd{J}_{  \bd{k} , \bd{Q}}^{(1 \pm)} .
\end{equation}
The contribution (\ref{eq:H1p1m})  to the Hamiltonian describes the 
hybridization between magnon and phonon modes. Note that for  finite $\bd{Q}$
the relevant coupling  $\bd{\Gamma}_{\bd{k}}^{Xb}$
vanishes linearly in $\bd{k}$
 for $\bd{k} \rightarrow 0$. 
As it will be discussed in Sec.~\ref{subsec:hybridization},
the hybridization term $H^{\mathrm{1pho}}_{\mathrm{1mag}}$ 
gives rise to an additional  renormalization of the phonon velocities 
which has the same order of magnitude as the
renormalization due to the contribution $\Sigma^{\mathrm{pho}}_0 ( \bd{k} , \lambda )$
arising from the classical spin background given in Eq.~(\ref{eq:sigmapho0}).

Note that the vertex $\bd{\Gamma}_{\bd{k}}^{Xb}$ does not have a definite symmetry
under $\bd{k} \rightarrow - \bd{k}$.
To obtain a more transparent classification of the
vertices, it is useful to  express the Holstein--Primakoff bosons
$b_{\bd{k}}$ in terms of Hermitian operators 
$\Phi_{\bd{k}} = \Phi^{\dagger}_{ - \bd{k}}$ and
$\Pi_{\bd{k}} = \Pi^{\dagger}_{ - \bd{k}}$ as follows:\cite{Hasselmann06,Kreisel08}
 \begin{align}
 b_{\bd{k}} & =  \sqrt{  \frac{ \Delta_{\bd{k}}}{2}} \Phi_{\bd{k}} +
 \frac{i}{\sqrt{2 \Delta_{\bd{k}}}} \Pi_{\bd{k}}, 
 \label{eq:hermitian1}
\\
 b^{\dagger}_{\bd{k}} & =  \sqrt{  \frac{ \Delta_{\bd{k}}}{2}} \Phi_{ - \bd{k}} -
 \frac{i}{\sqrt{2 \Delta_{\bd{k}}}} \Pi_{ - \bd{k}}, 
 \label{eq:hermitian2}
 \end{align}
where the energy $\Delta_{\bd{k}}$ is given by
 \begin{equation}
 \Delta_{\bd{k}} =  A_{\bd{k}}^{+} - B_{\bd{k}} =  A_{\bd{k}}^{+} + | B_{\bd{k}} |.
\end{equation}
Then our quadratic magnon Hamiltonian (\ref{eq:H2bb}) can be written as
 \begin{align}
 H_2 & = \frac{1}{2} \sum_{\bd{k}}  \Bigl\{ 
\Pi_{- \bd{k}} \Pi_{\bd{k}} 
 + \epsilon_{\bd{k}}^2 \Phi_{- \bd{k}} \Phi_{\bd{k}}
 \nonumber
 \\
 & 
+  i A_{\bd{k}}^- \left( \Phi_{- \bd{k}} \Pi_{\bd{k}} - \Phi_{ \bd{k}} \Pi_{-\bd{k}} \right)
 - A_{\bd{k}}^{+} \Bigr\},
 \label{eq:H2PiPhi}
 \end{align} 
while the magnon-phonon hybridization (\ref{eq:H1p1m}) reads
 \begin{align}
 H_{\mathrm{1mag}}^{\mathrm{1pho}}  & =   \frac{1}{2} \sum_{\bd{k}}  \Bigl\{ 
 \mathbf{\Gamma}_{\bd{k}}^{ X \Phi} \cdot ( \bd{X}_{ - \bd{k}} \Phi_{\bd{k}}
 +  \bd{X}_{  \bd{k}} \Phi_{- \bd{k}} )
 \nonumber
 \\
 &  \hspace{8mm}
+ \mathbf{\Gamma}_{\bd{k}}^{ X \Pi} \cdot ( \bd{X}_{ - \bd{k}} \Pi_{\bd{k}}
 -   \bd{X}_{  \bd{k}} \Pi_{- \bd{k}} ) \Bigr\}, \hspace{7mm}
 \label{eq:H11pp}
 \end{align}
where the vertex  $\mathbf{\Gamma}_{\bd{k}}^{ X \Phi}$
describing the coupling of the phonon coordinates to the generalized
magnon coordinates $\Phi_{\bd{k}}$ is an even function of $\bd{k}$,
while the corresponding vertex $\mathbf{\Gamma}_{\bd{k}}^{ X \Pi}$ 
which couples the phonon coordinates to the magnon momenta
is odd in $\bd{k}$:
 \begin{subequations}
 \begin{align}
  \mathbf{\Gamma}_{\bd{k}}^{ X \Phi} & =  \sqrt{\frac{  \Delta_{\bd{k}}}{2} }
 \left[ \mathbf{\Gamma}_{\bd{k}}^{ X b} + \mathbf{\Gamma}_{-\bd{k}}^{ X b}
 \right]  =   \mathbf{\Gamma}_{-\bd{k}}^{ X \Phi}  ,
 \\
 \mathbf{\Gamma}_{\bd{k}}^{ X \Pi} & =  \frac{i}{ \sqrt{2   \Delta_{\bd{k}} }}
 \left[ \mathbf{\Gamma}_{\bd{k}}^{ X b} - \mathbf{\Gamma}_{-\bd{k}}^{ X b}
 \right] =  -  \mathbf{\Gamma}_{-\bd{k}}^{ X \Pi}.
 \label{eq:GammaXPi}
\end{align}
 \end{subequations}
In principle, the complete quadratic part 
$H_2 + H^{\rm pho} + H_0^{\rm 2 pho} + H_{\rm 1 mag}^{\rm 1 pho}$
of the magnon-phonon Hamiltonian can be
diagonalized by means of a canonical transformation.\cite{Perelomov86}
But due to the absence of inversion symmetry of
the magnon dispersion $E_{\bd{k}}$
the explicit construction of this transformation seems to be a rather complicated
mathematical problem which we have not attempted to solve.
Fortunately, the renormalized phonon spectrum can be obtained in a straightforward 
manner at this level of approximation by
integrating in the path integral formulation of the theory
over the magnon field. As will be shown in Sec.~\ref{subsec:hybridization}, the
phonon spectrum can then be directly obtained from the inverse phonon propagator
appearing in the resulting Gaussian effective 
action for the phonon field.

\subsubsection{One-phonon two-magnon interaction}

The part of the spin-phonon Hamiltonian involving one phonon and two
magnon operators is
 \begin{align}
 H^{\mathrm{1pho}}_{\mathrm{2mag}} & =  \frac{S}{2} \sum_{ij} U^{(1)}_{ ij} 
 \Bigl\{ 
 -  \hat{\bd{m}}_i \cdot \hat{\bd{m}}_j  ( n_i + n_j ) 
 \nonumber
 \\
 & 
+
\frac{1}{2} \bigl[  (\bd{e}^+_i \cdot \bd{e}^-_j)  b^{\dagger}_i b_j  +    
(\bd{e}^-_i \cdot \bd{e}^+_j)   b_i  b_j^{\dagger}
 \nonumber
 \\
 &   \hspace{3mm} +  (\bd{e}^+_i \cdot \bd{e}^+_j)  b^{\dagger}_i b^{\dagger}_j 
 +   (\bd{e}^-_i \cdot \bd{e}^-_j)    b_i b_j 
 \bigr] \Bigr\}.
 \label{eq:H11}
 \end{align}
Using Eqs.~(\ref{eq:ei1}--\ref{eq:eip}) to calculate the relevant scalar products
of the basis vectors, we obtain after
Fourier transformation to momentum space,
  \begin{align}
 H^{\mathrm{1pho}}_{\mathrm{2mag}} & = 
\frac{1}{\sqrt{N}}
\sum_{ \bd{k} \bd{k}^{\prime}} 
 \biggl[ \bd{\Gamma}^{b^{\dagger}b}_{ \bd{k} , \bd{k}^{\prime}} \cdot
 \bd{X}_{ \bd{k} - \bd{k}^{\prime}} b^{\dagger}_{\bd{k}} b_{\bd{k}^{\prime}}
 \nonumber
 \\
 &  \hspace{-10mm}+ \frac{1}{ 2 !} \left( \bd{\Gamma}^{b^{\dagger} b^{\dagger}   }_{ \bd{k} , \bd{k}^{\prime}} \cdot
 \bd{X}_{ \bd{k} + \bd{k}^{\prime}} b^{\dagger}_{\bd{k}} b^{\dagger}_{\bd{k}^{\prime}}
 + 
\bd{\Gamma}^{ b b }_{ \bd{k} , \bd{k}^{\prime}} \cdot
 \bd{X}_{-\bd{k} - \bd{k}^{\prime}} b_{\bd{k}} b_{\bd{k}^{\prime}}
\right)
 \biggr],
 \label{eq:H1p2mfourier}
 \end{align}
where the coupling functions are
 \begin{align}
 \bd{\Gamma}^{     b^{\dagger} b }_{ \bd{k} , \bd{k}^{\prime}}
 & = 
\frac{S}{2} \biggl\{ ( 2 s_{\theta}^2 -  c_{\theta}^2 )  ( \bd{J}^{(1-)}_{\bd{k}, \bd{Q}} 
- 
\bd{J}^{(1-)}_{\bd{k}^{\prime}, \bd{Q}} ) 
\nonumber
 \\
&  \hspace{4mm} +
 2 \left[ \bd{J}^{(1)}_{ \bd{k} - \bd{k}^{\prime} } - ( \bd{J}^{(1)}_{\bd{k}} 
- \bd{J}^ {(1)}_{\bd{k}^{\prime}} ) 
 \right]
 \nonumber
 \\
 &   \hspace{4mm}
 - 2 c_{\theta}^2 \left[ \bd{J}^{(1-)}_{ \bd{k} - \bd{k}^{\prime}, \bd{Q} } - 
( \bd{J}^{(1-)}_{\bd{k}, \bd{Q}} -  \bd{J}^{(1-)}_{\bd{k}^{\prime} , \bd{Q}} ) \right]
 \nonumber
 \\
 &    \hspace{4mm}
+ 2 s_{\theta}  ( \bd{J}^{(1+)}_{\bd{k}, \bd{Q}} - 
\bd{J}^{(1+)}_{\bd{k}^{\prime}, \bd{Q}} )
\biggr\},
\label{eq:coupln}
 \\
\bd{\Gamma}^{b^{\dagger} b^{\dagger}   }_{ \bd{k} , \bd{k}^{\prime}} 
& = - \bd{\Gamma}^{ b b }_{ \bd{k} , \bd{k}^{\prime}}
= \frac{S}{2} c_{\theta}^2 \left[ \bd{J}^{(1-)}_{ \bd{k} , \bd{Q}} 
 +  \bd{J}^{(1-)}_{ \bd{k}^{\prime} , \bd{Q}}  \right].
 \label{eq:coupla}
 \end{align}
Since the Fourier transform $\bd{J}^{(1)}_{\bd{k}}$
of the derivative of the exchange coupling is purely imaginary,
this is also true for the above vertex functions
$\bd{\Gamma}^{     b^{\dagger} b }_{ \bd{k} , \bd{k}^{\prime}}$,
$\bd{\Gamma}^{b^{\dagger} b^{\dagger}   }_{ \bd{k} , \bd{k}^{\prime}} $ and $ \bd{\Gamma}^{ b b }_{ \bd{k} , \bd{k}^{\prime}}$.
Note also that the normal vertex
$\bd{\Gamma}^{     b^{\dagger} b }_{ \bd{k} , \bd{k}^{\prime}}$
is antisymmetric  while the anomalous vertices
$\bd{\Gamma}^{b^{\dagger} b^{\dagger}   }_{ \bd{k} , \bd{k}^{\prime}} $ and $ \bd{\Gamma}^{ b b }_{ \bd{k} , \bd{k}^{\prime}}$ are symmetric under $\bd{k} \leftrightarrow
 \bd{k}^{\prime}$. Keeping in mind that these vertices are purely imaginary,
one easily verifies the following relations,
 \begin{subequations}
 \begin{align}
\bd{\Gamma}^{b^{\dagger} b   }_{ \bd{k} , \bd{k}^{\prime}} & = 
 - \bd{\Gamma}^{b^{\dagger} b  }_{ \bd{k}^{\prime} , \bd{k}} =
 ( \bd{\Gamma}^{b^{\dagger} b  }_{ \bd{k}^{\prime} , \bd{k}})^{\ast} ,
 \label{eq:gbdaggerbsym} 
 \\
\bd{\Gamma}^{b^{\dagger} b^{\dagger}   }_{ \bd{k} , \bd{k}^{\prime}} & = 
\bd{\Gamma}^{b^{\dagger} b^{\dagger}   }_{ \bd{k}^{\prime} , \bd{k}}
=- \bd{\Gamma}^{b b   }_{ \bd{k}^{\prime} , \bd{k}}
= (\bd{\Gamma}^{b b   }_{ \bd{k}^{\prime} , \bd{k}})^{\ast},
 \label{eq:gbbdaggersym} 
\\
\bd{\Gamma}^{b b   }_{ \bd{k} , \bd{k}^{\prime}} & = 
 \bd{\Gamma}^{b b   }_{ \bd{k}^{\prime} , \bd{k}}
= - \bd{\Gamma}^{b^{\dagger} b^{\dagger}   }_{ \bd{k}^{\prime} , \bd{k}}
= (\bd{\Gamma}^{b^{\dagger} b^{\dagger}   }_{ \bd{k}^{\prime} , \bd{k}})^{\ast}.
 \label{eq:gbbsym}
 \end{align}
 \end{subequations}
Of particular interest is the
leading behavior  of the above magnon-phonon
vertices for small $\bd{k}$ and $\bd{k}^{\prime}$,
 \begin{align}
 \bd{\Gamma}^{b^{\dagger} b}_{ \bd{k} , \bd{k}^{\prime}} & = 
 \frac{S}{2} \biggl\{
 ( 2 s_{\theta}^2 - c_{\theta}^2 )  \bigl[
( \bd{k} - \bd{k}^{\prime}) \cdot \mathbf{\nabla}_{\bd{Q}} \bigr]
 \bigl[   \left. \bd{J}^{(1)}_{\bd{Q}} \right|_{\bd{Q}=0} -  \bd{J}^{(1)}_{\bd{Q}} 
\bigr] 
 \nonumber
 \\
 & \hspace{4mm}  -
  s_{\theta} \bigl[ ( \bd{k} \cdot \mathbf{\nabla}_{\bd{Q}} )^2
 -  ( \bd{k}^{\prime} \cdot \mathbf{\nabla}_{\bd{Q}} )^2 \bigr] 
\bd{J}^{(1)}_{\bd{Q}} 
\biggr\} +  {\mathcal{O}} ( \bd{k}^3),
 \nonumber
 \\
 & 
 \label{eq:Gammabdbdlong}
 \\
  \bd{\Gamma}^{b^{\dagger} b^{\dagger}}_{ \bd{k} , \bd{k}^{\prime}} & =
  \frac{S}{2} c_{\theta}^2  \bigl[  
( \bd{k} + \bd{k}^{\prime}) \cdot  \mathbf{\nabla}_{\bd{Q}} \bigr]
 \bigl[ \left. \bd{J}^{(1)}_{\bd{Q}} \right|_{\bd{Q}=0} -  \bd{J}^{(1)}_{\bd{Q}} 
\bigr] +  {\mathcal{O}} ( \bd{k}^3).
 \label{eq:Gammabblong}
 \end{align}
It turns out that $H^{\mathrm{1pho}}_{2\mathrm{mag}}$
in Eq.~(\ref{eq:H1p2mfourier}) is not the only part of the Hamiltonian describing
the coupling of a single phonon to two magnon operators.
The reason is that due to the magnon-phonon
hybridization discussed in Sec.~\ref{subsec:oneone} 
the Holstein--Primakoff bosons are a linear combination of true magnon operators and
phonon operators. In principle, 
the proper linear combination can be obtained by diagonalizing the quadratic 
magnon-phonon Hamiltonian 
$H_2 + H^{\mathrm{pho}}  + H_0^{\rm 2 pho} + H^{\mathrm{1pho}}_{\mathrm{1mag}}$ 
by means of a canonical
transformation.  Although the construction of such a transformation
is in principle possible,\cite{Perelomov86} for our purposes 
it is fortunately not necessary to explicitly solve this
complicated algebraic problem.
The reason is that
at long wavelengths and to the order in the small parameter $1/S$ 
consistent with Eq.~(\ref{eq:H11}), the proper magnon operators $\tilde{b}_{\bd{k}}$
can be determined from the requirement that in the quadratic part of the
magnon-phonon Hamiltonian
 $H_2 + H^{\mathrm{pho}} +  H_0^{\rm 2 pho}
+   H^{\mathrm{1pho}}_{\mathrm{1mag}}$ there should be no terms
describing the coupling of the  phonon coordinates $\bd{X}_{\bd{k}}$ 
to the generalized magnon momenta
${\Pi}_{\bd{k}}$, as in the second line of Eq.~(\ref{eq:H11pp}).
It is then easy to show that the proper magnon operators $\tilde{b}_{\bd{k}}$
are related to the original Holstein--Primakoff magnons $b_{\bd{k}}$ via the
phonon-dependent shift transformation 
 \begin{equation}
 {b}_{\bd{k}}  =  \tilde{b}_{\bd{k}} + \bd{\lambda}_{ \bd{k}}  \cdot \bd{X}_{\bd{k}},
 \label{eq:bshift}
 \end{equation}
where
 \begin{equation}
  \bd{\lambda}_{\bd{k}} = \frac{i}{\sqrt{2 \Delta_{\bd{k}}}} \bd{\Gamma}_{\bd{k}}^{X \Pi}
 \label{eq:lambdaknewdef}
 \end{equation}
depends on the antisymmetric vertex $\bd{\Gamma}_{\bd{k}}^{X \Pi}$
defined in Eq.~(\ref{eq:GammaXPi}).
If we now substitute the transformation (\ref{eq:bshift}) into 
the part $H_3$ of the pure magnon Hamiltonian 
involving three powers of the
Holstein--Primakoff bosons given in Eq.~(\ref{eq:H3hp}),
we generate (among other terms) an additional contribution to the
one-phonon two-magnon interaction,
which involves the same power of $S$ as
$H^{\mathrm{1pho}}_{\mathrm{2mag}}$ in Eq.~(\ref{eq:H11}) and therefore
should be taken into account on equal footing with 
$H^{\mathrm{1pho}}_{\mathrm{2mag}}$. Fortunately, 
this contribution can be absorbed via a simple redefinition
of the vertices in Eq.~(\ref{eq:H1p2mfourier}),
 \begin{subequations}
 \begin{align}
\tilde{\bd{\Gamma}}^{b^{\dagger }b}_{ \bd{k} , \bd{k}^{\prime}} & = 
 {\bd{\Gamma}}^{b^{\dagger }b}_{ \bd{k} , \bd{k}^{\prime}} +
\delta  {\bd{\Gamma}}^{b^{\dagger }b}_{ \bd{k} , \bd{k}^{\prime}} ,
 \label{eq:Gmp1}
 \\
\tilde{\bd{\Gamma}}^{b^{\dagger }b^{\dagger}}_{ \bd{k} , \bd{k}^{\prime}} & = 
 {\bd{\Gamma}}^{b^{\dagger }b^{\dagger}}_{ \bd{k} , \bd{k}^{\prime}} +
\delta  {\bd{\Gamma}}^{b^{\dagger }b^{\dagger}}_{ \bd{k} , \bd{k}^{\prime}} ,
 \label{eq:Gmp2}
 \\
\tilde{\bd{\Gamma}}^{bb }_{ \bd{k} , \bd{k}^{\prime}} & = 
 {\bd{\Gamma}}^{bb}_{ \bd{k} , \bd{k}^{\prime}} +
\delta  {\bd{\Gamma}}^{bb}_{ \bd{k} , \bd{k}^{\prime}} ,
 \label{eq:Gmp3}
 \end{align}
 \end{subequations}
where the correction terms due to magnon-phonon hybridization are
 \begin{subequations}
 \begin{align}
 \delta \bd{\Gamma}^{b^{\dagger}b}_{ \bd{k} , \bd{k}^{\prime}}
& =  \Gamma_3^{b^{\dagger} b^{\dagger} b } 
( \bd{k} - \bd{k}^{\prime} , - \bd{k} ; \bd{k}^{\prime} ) \bd{\lambda}_{ \bd{k}^{\prime} - \bd{k} }
 \nonumber
 \\
&+ 
  \Gamma_3^{b^{\dagger} b b } 
( - \bd{k} ; \bd{k}^{\prime} ,  \bd{k} - \bd{k}^{\prime} ) \bd{\lambda}_{ \bd{k} -\bd{k}^{\prime} },
 \label{eq:deltaGamma1}
 \\
 \delta \bd{\Gamma}^{b^{\dagger} b^{\dagger}}_{ \bd{k} , \bd{k}^{\prime}}
& =  \Gamma_3^{b^{\dagger} b^{\dagger} b } 
( - \bd{k} , -  \bd{k}^{\prime} ;   \bd{k} + \bd{k}^{\prime} ) 
\bd{\lambda}_{ \bd{k} + \bd{k}^{\prime}  },
 \label{eq:deltaGamma2}
 \\
 \delta \bd{\Gamma}^{b b}_{ \bd{k} , \bd{k}^{\prime}}
& =  \Gamma_3^{b^{\dagger} b b } 
( - \bd{k}  -  \bd{k}^{\prime} ;   \bd{k} , \bd{k}^{\prime} ) 
\bd{\lambda}_{ \bd{k} + \bd{k}^{\prime}  }.
 \label{eq:deltaGamma3}
 \end{align}
\end{subequations}
The effective one-phonon two-magnon Hamiltonian
$\tilde{H}^{\mathrm{1pho}}_{\mathrm{2mag}}$ 
can then be obtained from ${H}^{\mathrm{1pho}}_{\mathrm{2mag}}$
in Eq.~(\ref{eq:H1p2mfourier}) by substituting
the bare vertices by the shifted ones,
${\bd{\Gamma}}^{b^{\dagger }b}_{ \bd{k} , \bd{k}^{\prime}} \rightarrow
 \tilde{{\bd{\Gamma}}}^{b^{\dagger }b}_{ \bd{k} , \bd{k}^{\prime}} $,
${\bd{\Gamma}}^{b^{\dagger }b^{\dagger}}_{ \bd{k} , \bd{k}^{\prime}} \rightarrow
 \tilde{{\bd{\Gamma}}}^{b^{\dagger }b^{\dagger}}_{ \bd{k} , \bd{k}^{\prime}} $,
and
${\bd{\Gamma}}^{b b}_{ \bd{k} , \bd{k}^{\prime}} \rightarrow
 \tilde{{\bd{\Gamma}}}^{bb}_{ \bd{k} , \bd{k}^{\prime}} $.
Using the fact that according to Eqs.~(\ref{eq:Gammabbb1}--\ref{eq:K0lim})
at long wavelengths the three-magnon vertices
can be approximated by
 \begin{align}
  \Gamma_3^{b^{\dagger} b^{\dagger} b } ( \bd{k}_1 , \bd{k}_2 ; \bd{k}_3 )
  & =  -  \Gamma_3^{b^{\dagger} b b } ( - \bd{k}_3 , - \bd{k}_2 ; - \bd{k}_1 )
\nonumber
 \\
 &  =  
  - c_{\theta}   s_{\theta} \frac{ \sqrt{2S}}{i}   \frac{h_c}{S} +  {\mathcal{O}} ( \bd{k}^2 ),
 \end{align}
we  obtain for the shifts of the  one-phonon two-magnon vertices 
due to magnon-phonon hybridization at long wavelengths,
 \begin{align}
 \delta \bd{\Gamma}^{b^{\dagger} b}_{ \bd{k} , \bd{k}^{\prime}} & =
 - 2 S  s_{\theta}^2  \bigl[
( \bd{k} - \bd{k}^{\prime}) \cdot \mathbf{\nabla}_{\bd{Q}} \bigr]
 \bigl[  \left. \bd{J}^{(1)}_{\bd{Q}} \right|_{\bd{Q}=0} -  \bd{J}^{(1)}_{\bd{Q}} 
\bigr]  
 \nonumber
  \\
 &
+ {\mathcal{O}} ( \bd{k}^3 ),
 \label{eq:Gammashift1}
 \\
  \delta \bd{\Gamma}^{b^{\dagger} b^{\dagger}}_{ \bd{k} , \bd{k}^{\prime}} & = 
  S s_{\theta}^2  \bigl[  
( \bd{k} + \bd{k}^{\prime}) \cdot  \mathbf{\nabla}_{\bd{Q}} \bigr]
\bigl[  \left. \bd{J}^{(1)}_{\bd{Q}} \right|_{\bd{Q}=0} -  \bd{J}^{(1)}_{\bd{Q}} 
\bigr]
 \nonumber
 \\
 &
  +  {\mathcal{O}} ( \bd{k}^3 )    .
 \label{eq:Gammashift2}
 \end{align}
Combining these expressions with the long-wavelength limits
of the corresponding bare vertices given in
Eqs.~(\ref{eq:Gammabdbdlong}, \ref{eq:Gammabblong}), we find that the
proper one-phonon two-magnon vertices are
in the long-wavelength limit given by
 \begin{align}
 \tilde{\bd{\Gamma}}^{b^{\dagger} b}_{ \bd{k} , \bd{k}^{\prime}} & = 
 \nonumber
 \\
 &  \hspace{-8mm}
 - \frac{S}{2} \biggl\{
 ( 2 s_{\theta}^2 + c_{\theta}^2 )  \bigl[
( \bd{k} - \bd{k}^{\prime}) \cdot \mathbf{\nabla}_{\bd{Q}} \bigr]
 \bigl[  \left. \bd{J}^{(1)}_{\bd{Q}} \right|_{\bd{Q}=0} -  \bd{J}^{(1)}_{\bd{Q}} 
\bigr]  
 \nonumber
 \\
 &  \hspace{-5mm}
+  s_{\theta} \bigl[ ( \bd{k} \cdot \mathbf{\nabla}_{\bd{Q}} )^2
 -  ( \bd{k}^{\prime} \cdot \mathbf{\nabla}_{\bd{Q}} )^2 \bigr] 
\bd{J}^{(1)}_{\bd{Q}} 
\biggr\} + {\mathcal{O}} ( \bd{k}^3 ),
 \label{eq:Gamma21n}
 \\
  \tilde{\bd{\Gamma}}^{b^{\dagger} b^{\dagger}}_{ \bd{k} , \bd{k}^{\prime}} & = 
  \frac{S}{2} ( 2 s_{\theta}^2 + c_{\theta}^2)  \bigl[  
( \bd{k} + \bd{k}^{\prime}) \cdot  \mathbf{\nabla}_{\bd{Q}} \bigr]
 \bigl[ \left. \bd{J}^{(1)}_{\bd{Q}} \right|_{\bd{Q}=0} -  \bd{J}^{(1)}_{\bd{Q}} 
\bigr] 
\nonumber
 \\
 & 
+  {\mathcal{O}} ( \bd{k}^3 )    .
 \label{eq:Gamma21a}
 \end{align}
Note that the coefficients of the terms linear in the momenta 
differ only by a minus sign, which will turn out to be essential
to obtain the correct long-wavelength limit of the
ultrasonic attenuation rate.

At this point 
it is convenient to express 
the effective one-phonon two-magnon Hamiltonian
$\tilde{H}^{\mathrm{1pho}}_{\mathrm{2mag}}$ 
in terms of the Bogoliubov quasiparticle
operators $\beta_{\bd{k}}$ and $\beta^{\dagger}_{\bd{k}}$
defined in Eq.~(\ref{eq:Bogoliubov}).
We obtain
   \begin{align}
 \tilde{H}^{\mathrm{1pho}}_{\mathrm{2mag}} & = 
\frac{1}{\sqrt{N}}
\sum_{ \bd{k} \bd{k}^{\prime}} 
 \biggl[ \tilde{\bd{\Gamma}}^{\beta^{\dagger}\beta}_{ \bd{k} , \bd{k}^{\prime}} \cdot
 \bd{X}_{ \bd{k} - \bd{k}^{\prime}} \beta^{\dagger}_{\bd{k}}  \beta_{\bd{k}^{\prime}}
 \nonumber
 \\
 &  \hspace{-10mm}+ \frac{1}{ 2 !} \left( \tilde{\bd{\Gamma}}^{\beta^{\dagger} \beta^{\dagger}   }_{ \bd{k} , \bd{k}^{\prime}} \cdot
 \bd{X}_{ \bd{k} + \bd{k}^{\prime}} \beta^{\dagger}_{\bd{k}} \beta^{\dagger}_{\bd{k}^{\prime}}
 + 
\tilde{\bd{\Gamma}}^{ \beta \beta }_{ \bd{k} , \bd{k}^{\prime}} \cdot
 \bd{X}_{ -\bd{k} - \bd{k}^{\prime}} \beta_{\bd{k}} \beta_{\bd{k}^{\prime}}
\right)
 \biggr],
 \label{eq:H1p2mbeta}
 \end{align}
where the vertices are given by
\begin{subequations}
 \begin{align}
\tilde{\bd{\Gamma}}^{\beta^{\dagger}\beta}_{ \bd{k} , \bd{k}^{\prime}} & = 
u_{\bd{k}} u_{\bd{k}^{\prime}}
\tilde{\bd{\Gamma}}^{b^{\dagger}b }_{ \bd{k} , \bd{k}^{\prime}}
- v_{\bd{k}} v_{\bd{k}^{\prime}}
\tilde{\bd{\Gamma}}^{b^{\dagger} b }_{ - \bd{k} , - \bd{k}^{\prime}}
 \nonumber
 \\
 & - 
   u_{\bd{k}} v_{\bd{k}^{\prime}}
\tilde{\bd{\Gamma}}^{b^{\dagger}b^{\dagger} }_{ \bd{k} , - \bd{k}^{\prime}}
- v_{\bd{k}} u_{\bd{k}^{\prime}}
\tilde{\bd{\Gamma}}^{bb }_{ - \bd{k} ,  \bd{k}^{\prime}} ,
 \label{eq:Gammabb1}
 \\
\tilde{\bd{\Gamma}}^{ \beta^{\dagger} \beta^{\dagger}   }_{ \bd{k} , \bd{k}^{\prime}}
& = 
 u_{\bd{k}} u_{\bd{k}^{\prime}} 
\tilde{\bd{\Gamma}}^{b^{\dagger} b^{\dagger}   }_{ \bd{k} , \bd{k}^{\prime}}
+ v_{\bd{k}} v_{\bd{k}^{\prime}}
\tilde{\bd{\Gamma}}^{b b   }_{ - \bd{k} , - \bd{k}^{\prime}}
 \nonumber
 \\
 & - 
  u_{\bd{k}} v_{\bd{k}^{\prime}} \tilde{\bd{\Gamma}}^{b^{\dagger}b }_{  \bd{k} ,  -\bd{k}^{\prime}}
+  v_{\bd{k}} u_{\bd{k}^{\prime}} \tilde{\bd{\Gamma}}^{b^{\dagger}b }_{  - \bd{k}, 
\bd{k}^{\prime} }
 , 
 \label{eq:Gammabb2}
\\
 \tilde{\bd{\Gamma}}^{\beta \beta   }_{ \bd{k} , \bd{k}^{\prime}} & = 
 u_{\bd{k}} u_{\bd{k}^{\prime}} 
\tilde{\bd{\Gamma}}^{b b   }_{ \bd{k} , \bd{k}^{\prime}}
+ v_{\bd{k}} v_{\bd{k}^{\prime}}
\tilde{\bd{\Gamma}}^{b^{\dagger} b^{\dagger}   }_{ - \bd{k} , - \bd{k}^{\prime}}
 \nonumber
 \\
 & +   u_{\bd{k}} v_{\bd{k}^{\prime}} \tilde{\bd{\Gamma}}^{b^{\dagger}b }_{  \bd{k}, 
-\bd{k}^{\prime} }
-  v_{\bd{k}} u_{\bd{k}^{\prime}} 
\tilde{\bd{\Gamma}}^{b^{\dagger}b }_{  -\bd{k} ,  \bd{k}^{\prime}}.
 \label{eq:Gammabb3}
 \end{align}
\end{subequations}
As will be shown in Sec.~\ref{subsec:ultrasound}, 
at zero temperature the ultrasonic attenuation rate  is determined by the
anomalous vertex 
$\tilde{\bd{\Gamma}}^{ \beta^{\dagger} \beta^{\dagger}   }_{ \bd{k} , \bd{k}^{\prime}}$,
whose long-wavelength limit is explicitly given by
 \begin{align}
\tilde{\bd{\Gamma}}^{ \beta^{\dagger} \beta^{\dagger}   }_{ \bd{k} , \bd{k}^{\prime}}
 & =  \frac{S}{2} \biggl\{
 \nonumber
 \\
 & \hspace{-10mm}
 \left( \frac{2 s_{\theta}^2}{c_{\theta}^2} + 1 \right) 
\sqrt{\frac{ \epsilon_{\bd{k}} \epsilon_{\bd{k}^{\prime}}}{h_c^2} }
  \bigl[
( \bd{k} + \bd{k}^{\prime}) \cdot \mathbf{\nabla}_{\bd{Q}} \bigr]
 \bigl[  \left. \bd{J}^{(1)}_{\bd{Q}} \right|_{\bd{Q}=0} -  \bd{J}^{(1)}_{\bd{Q}} 
\bigr]  
 \nonumber
 \\
 & \hspace{-10mm}
+  \frac{s_{\theta}}{2} \frac{ \epsilon_{\bd{k}} - \epsilon_{\bd{k}^{\prime}}}{
 \sqrt{ \epsilon_{\bd{k}} \epsilon_{\bd{k}^{\prime}} } }
 \bigl[ ( \bd{k} \cdot \mathbf{\nabla}_{\bd{Q}} )^2
 -  ( \bd{k}^{\prime} \cdot \mathbf{\nabla}_{\bd{Q}} )^2 \bigr] 
\bd{J}^{(1)}_{\bd{Q}} 
\biggr\} + {\mathcal{O}} ( \bd{k}^3 ).
 \label{eq:Gammabebelong}
\end{align}

\section{Phonon self-energy due to magnon-phonon interactions}
\label{sec:renormalization}

\subsection{Elastic constants}
\label{subsec:hybridization}
The elastic constants are directly related to the velocities of the acoustic phonons 
which can be determined experimentally with high accuracy.\cite{Luethi} 
Although the spin-phonon coupling is expected to be small, its influence on the phonon 
properties is visible in the magnetic-field dependence
of the elastic constants.
The leading contributions to the  shift in the elastic constants
is already contained in the quadratic magnon-phonon Hamiltonian $ H_2 + H^{\mathrm{pho}} +   H^{\mathrm{2pho}}_0 +    H^{\mathrm{1pho}}_{\mathrm{1mag}}$. 
The phonon self-energy $\Sigma^{\mathrm{pho}}_0 ( \bd{k} , \lambda )$
given in Eq.~(\ref{eq:sigmapho0})
which is due to the coupling of the phonons to the classical spin background
renormalizes the phonon  frequencies according to
$\omega^2_{\bd{k} \lambda} \rightarrow \omega^2_{\bd{k} \lambda} +
\Sigma^{\mathrm{pho}}_0 ( \bd{k} , \lambda )$.
The self-energy correction leads to the following relative shift of the phonon velocities,
\begin{equation}
  \frac{ (\Delta c_{\lambda})_{0} }{c_{\lambda}} 
  =  \sqrt{1-  \lim_{ | \bd{k} | \rightarrow 0}    \frac{ \Sigma^{\mathrm{pho}}_0 ( \bd{k} , \lambda )    }{  
 \omega^2_{\bd{k} \lambda}   }}-1\;.
 \label{eq:Deltac0}
\end{equation}
An additional  renormalization of the phonon velocities arises from the
 magnon-phonon hybridization in \Ref{eq:H1p1m}.
To calculate this contribution, we note that
the phonons couple to the magnons only via the combination
$\bd{X}_{\bd{k}} = \sum_\lambda X_{\bd{k} \lambda} \bd{e}_{\bd{k} \lambda}$ 
so that it is convenient
to describe the phonon dynamics within an effective Lagrangian which is obtained
by integrating over the canonical phonon momenta $P_{ \bd{k} \lambda}$ in the
phase space functional integral representation of the theory.~\cite{Kreisel08} 
To leading order for large $S$ we may 
truncate  the effective Euclidean action at the
quadratic order in the fluctuations, 
\begin{equation}
 S [ \bd{X} , \bar{\beta} , \beta ] \approx S^{\mathrm{2pho}} [ \bd{X} ] + 
S_{\mathrm{2mag}} [ \bar{\beta} , \beta ]
 + S^{\mathrm{1pho}}_{\mathrm{1mag}} [ \bd{X} , \bar{\beta} , \beta ]     ,
 \end{equation}
where the Gaussian actions $S^{\mathrm{2pho}} [ \bd{X} ]$ and
 $S_{\mathrm{2mag}} [ \bar{\beta} , \beta ]$ describe noninteracting phonons and magnons,
\begin{align}
  S^{\mathrm{2pho}} [ \bd{X} ] & = \frac{1}{2 T} \sum_{ K  \lambda} 
M [ \omega^2 + \omega_{\bd{k} \lambda}^2 +
\Sigma^{\mathrm{pho}}_0 ( \bd{k} , \lambda )
] 
 \nonumber
 \\
& \hspace{15mm} \times 
X_{- K \lambda} X_{ K \lambda}\;,
 \label{eq:S2pho}
 \\
 S_{\mathrm{2mag}} [ \bar{\beta} , \beta ] & =  - \frac{1}{T} \sum_K ( i \omega - E_{\bd{k}} ) 
 \bar{\beta}_K \beta_K\;.
 \label{eq:S2mag}
 \end{align}
Here $T$ is the temperature and  $K = ( i \omega , \bd{k} )$ 
is a collective label containing
bosonic Matsubara frequencies $i \omega$ and wavevectors $\bd{k}$.
The real field $X_{K \lambda}$ represents  the Fourier components of the
phonon operator $X_{\bd{k} \lambda}$, while the complex boson field
$\beta_K$ represents the Fourier components of the magnon 
operator $\beta_{\bd{k}}$.
From Eqs.~(\ref{eq:Bogoliubov}, \ref{eq:H1p1m}) we see that the
magnon-phonon hybridization is represented by the Euclidean action
 \begin{align}
 S^{\mathrm{1pho}}_{\mathrm{1mag}} [ \bd{X} , \bar{\beta} , \beta ] & = \frac{1}{T} \sum_K
  \bd{\Gamma}^{X \beta}_{\bd{k}} \cdot \left( \bd{X}_{- K} \beta_K
+ \bd{X}_{K} \bar{\beta}_K \right), 
 \end{align}
where the hybridization vertex is the following linear combination
of the corresponding hybridization vertex $\bd{\Gamma}^{Xb}_{\bd{k}}$
in the Holstein--Primakoff basis given in Eq.~(\ref{eq:Gamma1p1m}),
 \begin{equation}
 {\bd{\Gamma}}^{X \beta}_{\bd{k}} = u_{\bd{k}} \bd{\Gamma}^{Xb}_{\bd{k}}
 - v_{\bd{k}} \bd{\Gamma}^{Xb}_{-\bd{k}}.
 \label{eq:Gamma1pbeta}
\end{equation}
The Gaussian integral over the magnon field is now easily carried out,
and we obtain for the effective phonon action in Gaussian approximation,
 \begin{align}
S^{\mathrm{2pho}}_{\mathrm{eff}} [ \bd{X} ]  & =  \frac{1}{2 T} \sum_{ K  \lambda \lambda^{\prime}} 
 \biggl[  \delta_{\lambda , \lambda^{\prime}}
M [ \omega^2 + \omega_{\bd{k} \lambda}^2 + \Sigma^{\mathrm{pho}}_0 ( \bd{k} , \lambda )    ]
\nonumber
 \\
 &  \hspace{-7mm}
+ \frac{  
\bigl( {\bd{\Gamma}}^{X \beta}_{\bd{k}} \cdot \bd{e}_{\bd{k} \lambda} \bigr)^{\ast}
\bigl( {\bd{\Gamma}}^{X \beta}_{\bd{k}} \cdot \bd{e}_{\bd{k} \lambda^{\prime}} \bigr)
 }{
 i \omega - E_{\bd{k}}} \biggr]
X_{- K \lambda} X_{ K \lambda^{\prime}} .
\end{align}
For simplicity,  we neglect in the sum over the phonon modes
the optical phonons and  
off-diagonal terms $ \lambda \neq \lambda^{\prime}$.
In this approximation the magnon-phonon hybridization 
generates the following phonon self-energy:
 \begin{equation}
 \Sigma^{\mathrm{pho}}_{1} ( K , \lambda  )
 = \frac{  
\bigl|{\bd{\Gamma}}^{X \beta}_{\bd{k}} \cdot \bd{e}_{\bd{k} \lambda} \bigr|^2
 }{M ( 
 i \omega - E_{\bd{k}} ) }.
 \label{eq:sigmapho1}
\end{equation}
Combining this with the classical self-energy  $\Sigma^{\mathrm{pho}}_{0} ( {\bd{k}} , \lambda  )$
given in Eq.~(\ref{eq:sigmapho0}), we conclude that
the renormalized phonon dispersion $\tilde{\omega}_{ \bd{k} \lambda}$
is determined by the real positive root of the cubic equation
 \begin{align}
\tilde{\omega}_{ \bd{k} \lambda}^2 &= 
{\omega}_{ \bd{k} \lambda}^2+
\Sigma^{\mathrm{pho}}_{0} ( {\bd{k}} , \lambda  ) 
+ \Sigma^{\mathrm{pho}}_{1} ( i \omega \rightarrow \tilde{\omega}_{\bd{k} \lambda},
 \bd{k} ,  \lambda  )
 \notag \\
&={\omega}_{ \bd{k} \lambda}^2+ \Sigma^{\mathrm{pho}}_{0} ( \bd{k} , \lambda  )
 + \frac{  
\bigl|{\bd{\Gamma}}^{X \beta}_{\bd{k}} \cdot \bd{e}_{\bd{k} \lambda} \bigr|^2
 }{
 M( \tilde{\omega}_{\bd{k} \lambda}  - E_{\bd{k}} )}.
 \label{eq:omegatot}
 \end{align}
Recall that according to Eq.~(\ref{eq:sigmapho0}) the classical self-energy
$\Sigma^{\mathrm{pho}}_{0} ( \bd{k} , \lambda  )$ is proportional to $S^2$.
Keeping in mind that according to Eq.~(\ref{eq:Gamma1p1m}) the
hybridization vertex  ${\bd{\Gamma}}^{X \beta}_{\bd{k}}$
is of order $S^{3/2}$ and the denominator in Eq.~(\ref{eq:sigmapho1})
contains one power of $E_{\bd{k}} \propto S$, we see that
both self-energy contributions on the right-hand side of
Eq.~(\ref{eq:omegatot}) have the same order of magnitude
in a formal $1/S$-expansion.

Because in the experimentally relevant regime the magnon velocity
is small compared with the phonon velocity, the shift in
the phonon velocities
due to magnon-phonon hybridization can be approximated by
\begin{equation}
 \frac{ (\Delta c_{\lambda})_1}{c_{\lambda}} =
\lim_{ | \bd{k} | \rightarrow 0 } \frac{  
\bigl|{\bd{\Gamma}}^{X \beta}_{\bd{k}} \cdot \bd{e}_{\bd{k} \lambda} \bigr|^2
 }{
 2 M {\omega}_{\bd{k} \lambda}^3}\;.
 \label{eq:deltac}
 \end{equation}
Note that this contribution is always positive,
i.e. the coupling to the magnons  enhances the  phonon velocities, which
is a simple consequence of the fact that for Cs$_2$CuCl$_4$
the magnon velocities are small compared with the phonon velocities.
To explicitly take the limit in Eq.~(\ref{eq:deltac})
we need the small-momentum limit of the vertex
${\bd{\Gamma}}^{X \beta}_{\bd{k}}$. Using Eqs.~(\ref{eq:Gamma1p1m})
and (\ref{eq:Gamma1pbeta}) we obtain to leading order
  \begin{align}
 {\bd{\Gamma}}^{X \beta}_{\bd{k}}
 & =   \frac{i}{4} ( 2 S)^{3/2}   | \bd{k} |^{3/2} \sqrt{ \frac{   v(  \hat{\bd{k}} )  }{h_c} }   
{\bd{F}}^{X \beta} ( \hat{\bd{k}}  ),
\end{align}
where the dimensionless vector ${\bd{F}}^{X \beta} ( \hat{\bd{k}}  )$ is given by
 \begin{equation}
\bd{F}^{X \beta} ( \hat{\bd{k}}  ) =  s_{\vt} {\bd f}^{X\beta}_1(\hat{\bd k})-c_\vt^2 {\bd f}^{X\beta}_2 
(\hat{\bd k},\hat{\bd k})\;, \label{eq:Gammahybsmall}
\end{equation}
and we introduced the auxiliary vector functions
\begin{subequations}
\begin{align}
 {\bd f}^{X\beta}_1(\hat{\bd k})&=\frac{1}{h_c} 
 ( \hat{\bd{k}} \cdot \nabla_{\bd{Q}} ) 
  \Bigl[ \left.  {\bd{J}}^{(1)}_{\bd{Q}} \right|_{\bd{Q}=0}-{\bd{J}}^{(1)}_{\bd{Q}} \Bigr]\;,\label{eq:f1xb}\\
 {\bd f}^{X\beta}_2 (\hat{\bd k},\hat{\bd k}^\prime)&=
 \frac{1}{2 v ( \hat{\bd{k}} ) }
 ( \hat{\bd{k}} \cdot \nabla_{\bd{Q}} ) ( \hat{\bd{k}^\prime} \cdot \nabla_{\bd{Q}} ) 
 {\bd{J}}^{(1)}_{\bd{Q}}\;.\label{eq:f2xb}
 \end{align}
\end{subequations}
Here 
we have approximated the magnon dispersion by 
its leading long-wavelength limit,
 $
E_{\bd{k}} \approx \epsilon_{\bd{k}} \approx v ( \hat{\bd{k}} ) | \bd{k} |$;
see Eqs.~(\ref{eq:Eksmall}) and (\ref{eq:vkhat}).
Substituting Eq.~(\ref{eq:Gammahybsmall})
into Eq.~(\ref{eq:deltac}) we finally obtain
\begin{align}
 \frac{ (\Delta c_{\lambda})_1 }{c_{\lambda}} & =
 \frac{S^3}{4} \left( \frac{ v ( \hat{\bd{k}} )}{c_{\lambda} } \right)
 \left( \frac{h_c}{ M c_{\lambda}^2 } \right)
 \bigl| {\bd{F}}^{X \beta} ( \hat{\bd{k}} ) \cdot 
 \bd{e}_{\bd{k} \lambda} \bigr|^2
\notag
 \\
& =\frac{S^3}{4} \left( \frac{ v ( \hat{\bd{k}} )}{c_{\lambda} } \right)
 \left( \frac{h_c}{ M c_{\lambda}^2 } \right)
\notag
 \\
 & \hspace{7mm} \times
 \bigl| s_{\theta}
 {\bd f}^{X\beta}_1(\hat{\bd k}) \cdot  \bd{e}_{\bd{k} \lambda} 
- c_{\theta}^2  {\bd f}^{X\beta}_2  (\hat{\bd k},\hat{\bd k}) \cdot 
\bd{e}_{\bd{k} \lambda} 
\bigr|^2.
 \label{eq:deltac1}
 \end{align}
Formally, this is the leading-order contribution in the small parameters $v(\hat{\bd{k}})/c_\lambda$ and $h_c/(Mc_\lambda^2)$. 
Note that the magnetic-field dependence of the vertex (\ref{eq:Gammahybsmall}) 
is hidden in the canting angle $\theta$, which we approximate by its classical 
value (\ref{eq:tiltclassical}). 
Adding the classical and the hybridization contributions,
we finally obtain for the total velocity shift,
\begin{equation}
 \frac{ \Delta c_{\lambda}}{c_{\lambda}} =\frac{ (\Delta c_{\lambda} )_0+ 
(\Delta c_{\lambda})_1}{c_{\lambda}}\;.
\label{eq:deltacfull}
\end{equation}
Depending on the polarization vector ${\bd{e}}_{{\bd k} \lambda}$   of the
phonon and on the structure of the
first-derivative vector $\bd {J}^{(1)}_{  \bd{k}} $ 
and the second derivative tensor $\mathbf{J}^{(2)}_{\bd{k}}$,
either the first or the second contribution on the right-hand side of  
Eq.~(\ref{eq:deltacfull}) can dominate.
In Sec.~\ref{sec:comparison} we shall explicitly evaluate 
Eq.~(\ref{eq:deltacfull}) for the specific parameters of 
Cs$_2$CuCl$_4$ and compare our results with experiments.

\subsection{Ultrasonic attenuation}
\label{subsec:ultrasound}

\begin{figure}[tb]
  \includegraphics[width=0.48\textwidth]{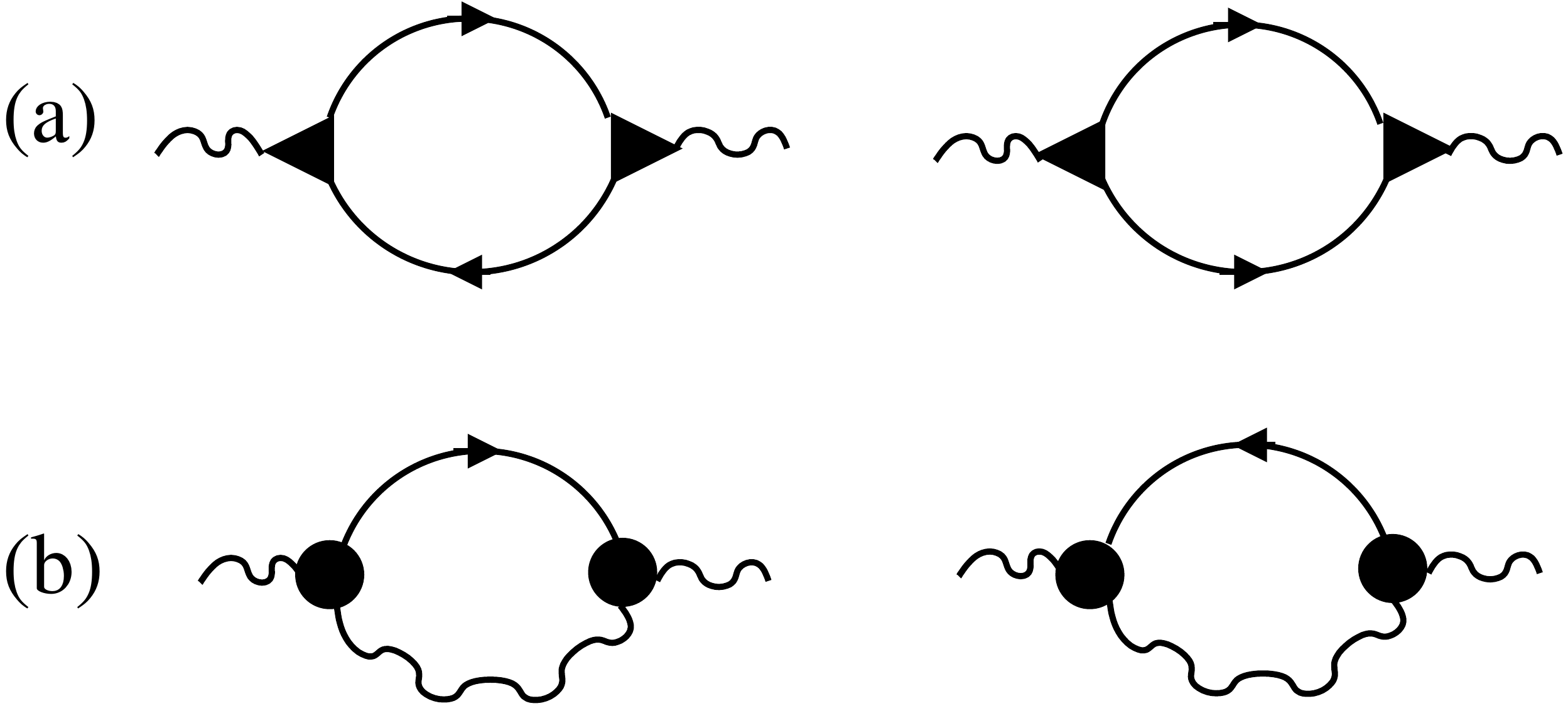}\centering
 \caption{These Feynman diagrams represent contributions to the  phonon self-energy
which determine the 
ultrasonic attenuation rate to leading order in the large-$S$ expansion.
(a) One-phonon two-magnon processes. (b) Two-phonon one-magnon processes.
Wavy lines represent the phonon Green's functions $G^{\rm pho} ( K \lambda )$,
solid arrows represent the magnon propagators $G_{\rm mag} ( K )$,
black triangles represent the shifted one-phonon two-magnon vertices 
defined in Eqs.~(\ref{eq:Gammabb1}--\ref{eq:Gammabb3}), while black circles
represent the two-phonon one-magnon vertices
defined in Eqs.~(\ref{eq:GammaXXbeta1}, \ref{eq:GammaXXbeta2}).
Both processes lead to an attenuation rate  proportional to $\bd{k}^4$ 
at long wavelengths.
The processes (a) exhibit a stronger dependence on the external magnetic field and
dominate if the magnon velocities are small compared with the phonon velocities.
}
 \label{fig:Feynman}
 \end{figure}

To obtain the magnetic-field dependence of the
ultrasonic attenuation rate, we should calculate the damping 
of the phonons due to the coupling to the magnon system.
In principle, the
attenuation rate can be obtained using Fermi's golden rule. 
Due to the rather complicated matrix elements of the relevant
interaction vertices, we find it more convenient to use a  many-body approach where
the damping rate is obtained from the imaginary part of the phonon self-energy.
The lowest-order interaction processes leading to phonon damping
are shown diagrammatically in Fig.~\ref{fig:Feynman}.
It turns out that only the processes in Fig.~\ref{fig:Feynman}(a) involving two magnons in the intermediate
lead to a strong magnetic-field dependence of the ultrasonic attenuation rate,
so that we shall focus here on these processes.
In the appendix we shall show that for 
Cs$_2$CuCl$_4$ where the magnon velocities are
small compared with the phonon velocities the other processes in Fig.~\ref{fig:Feynman}(b) 
involving one magnon and one phonon in the intermediate state
can indeed be neglected.
To obtain the contribution of the processes shown
in Fig.~\ref{fig:Feynman}(a) to the ultrasonic attenuation rate,
it is sufficient to retain only the effective one-phonon two-magnon interaction
process in our coupled magnon-phonon action. We thus approximate
 \begin{equation}
 S [ \bd{X} , \bar{\beta} , \beta ] = S^{\mathrm{2pho}} [ \bd{X} ] 
+ S_{ \mathrm{2mag}} [ \bar{\beta} , \beta ]
 + \tilde{S}_{ 2\mathrm{mag}}^{1\mathrm{pho}} [ \bd{X} , \bar{\beta} , \beta ]     ,
 \end{equation}
where the Gaussian parts  $S^{\mathrm{2pho}} [ \bd{X} ] $ and
$ S_{ \mathrm{2mag}} [ \bar{\beta} , \beta ]$ are given in 
Eqs.~(\ref{eq:S2pho}, \ref{eq:S2mag}), while the effective
magnon-phonon interaction is 
 \begin{align}
\tilde{S}^{\mathrm{1pho}}_{\mathrm{2mag}} [ \bd{X} , \bar{\beta} , \beta ] & = 
 \frac{ 1}{ \sqrt{N} T } \sum_{ K , K^{\prime}}
 \biggl[
 \tilde{\bd{\Gamma}}^{\beta^{\dagger}\beta}_{ \bd{k} , \bd{k}^{\prime}} \cdot
  \bd{X}_{ K - K^{\prime} }
 \bar{\beta}_{K}  {\beta}_{K^{\prime}} 
 \nonumber
 \\
 &  \hspace{-25mm}+ \frac{1}{ 2 !}  \left(
 \tilde{\bd{\Gamma}}^{\beta^{\dagger} \beta^{\dagger}   }_{ \bd{k} , \bd{k}^{\prime}} 
\cdot   \bd{X}_{K + K^{\prime} }
\bar{\beta}_K \bar{\beta}_{K^{\prime}}
 + \tilde{\bd{\Gamma}}^{ \beta \beta }_{ \bd{k} , \bd{k}^{\prime}} 
 \cdot  \bd{X}_{ - K - K^{\prime} }
\beta_K \beta_{K^{\prime}}  \right)
 \biggr],
 \label{eq:S1p2m}
 \end{align}
which corresponds to the effective one-phonon two-magnon Hamiltonian
$\tilde{H}^{\mathrm{1pho}}_{\mathrm{2mag}}$ defined in Eq.~(\ref{eq:H1p2mbeta}).
Note that the vertices in this expression
take the magnon-phonon hybridization
into account. The phonon self-energy
$\Sigma^{\rm pho} ( K  \lambda )$ and the magnon self-energy
$\Sigma_{\rm mag} ( K )$ 
are defined in terms of the corresponding Green's functions:
 \begin{align}
 G^{\rm pho} ( K \lambda ) & =  \frac{ M}{T}  \langle X_{ - K \lambda} X_{ K \lambda} 
 \rangle  =
\frac{1}{  \omega^2 + \omega_{\bd{k} \lambda}^2 
+ \Sigma^{\rm pho} ( K \lambda ) },
  \\
 G_{\rm mag} ( K  ) & =  - \frac{1}{T}   \langle 
\bar{\beta}_{  K } \beta_{ K } 
 \rangle  = \frac{1}{  i \omega   -E_{\bd{k}} - \Sigma_{ \rm mag} ( K ) }.
 \end{align}
The diagrams shown in Fig.~\ref{fig:Feynman} (a) give rise to the following contribution
to the phonon self-energy,
 \begin{align}
   \Sigma^{\rm pho}_2 ( K \lambda ) & =  - \frac{T}{ N} \sum_{ K^{\prime}}
 \biggl[ 
 \nonumber
 \\
 &  \hspace{-10mm}
\frac{  | \tilde{\bd{\Gamma}}^{\beta^{\dagger} \beta}_{ \bd{k}^{\prime} , 
 \bd{k}^{\prime} - \bd{k} } \cdot \bd{e}_{ \bd{k} \lambda} |^2}{2M}
 G_{ \rm mag} ( K^{\prime} )
G_{\rm  mag} ( K^{\prime} -K )
 \nonumber
 \\
&  \hspace{-15mm}  +
\frac{  | \tilde{\bd{\Gamma}}^{\beta^{\dagger} \beta^{\dagger}}_{ \bd{k}^{\prime} , 
 \bd{k} - \bd{k}^{\prime}  } \cdot \bd{e}_{ \bd{k} \lambda} |^2}{2M}
 G_{\rm mag} ( K^{\prime} )
G_{\rm mag} ( K- K^{\prime}  ) 
 \biggr] 
 \nonumber
 \\
&  \hspace{-15mm}
+ ( K \rightarrow - K )  .
 \label{eq:selfX}
 \end{align}
Neglecting self-energy corrections to the magnon propagators,
 the frequency summation can be easily carried out,
\begin{align}
&   \Sigma^{\rm pho}_{2} ( K \lambda )  = 
\nonumber
 \\
 & 
 -  \frac{1}{ N} \sum_{ \bd{k}^{\prime}}
 \biggl[  
  \frac{| \tilde{\bd{\Gamma}}^{\beta^{\dagger} \beta}_{ \bd{k}^{\prime} , 
 \bd{k}^{\prime} - \bd{k} } \cdot \bd{e}_{ \bd{k} \lambda} |^2}{2M}
\frac{ n ( E_{\bd{k}^{\prime}} ) -   n ( E_{\bd{k}^{\prime} - \bd{k}} )}{ i \omega 
 - E_{\bd{k}^{\prime}} + E_{\bd{k}^{\prime} - \bd{k}} }
 \nonumber
 \\
&  +  
 \frac{
  | \tilde{\bd{\Gamma}}^{\beta^{\dagger} \beta^{\dagger}}_{ \bd{k}^{\prime} , 
 \bd{k} - \bd{k}^{\prime}  } \cdot \bd{e}_{ \bd{k} \lambda} |^2}{2M}
 \frac{ n ( E_{\bd{k}^{\prime}} ) +   n ( E_{    \bd{k} - \bd{k}^{\prime} } ) + 1}{ - i \omega 
 + E_{\bd{k}^{\prime}} + E_{   \bd{k} - \bd{k}^{\prime} } } \biggr]
  \nonumber
 \\
 & 
+ ( K \rightarrow - K ),
 \label{eq:selfXmat0}
 \end{align}
where $n ( E ) = 1/( e^{ E/T} -1 )$ is the Bose function.
To calculate the phonon damping, we analytically continue this expression to the
real frequency axis, $i \omega \rightarrow \omega + i 0$, and take the imaginary part.
We obtain
 \begin{widetext}
\begin{align}
  {\mathrm{Im}} \Sigma^{\rm pho}_{2} ( \omega + i0, \bd{k} ,  \lambda ) & =  -   [ 1 - e^{ -  \omega/T }]
\frac{\pi}{ N} \sum_{ \bd{k}^{\prime}}
 \biggl[ 
  \frac{| \tilde{\bd{\Gamma}}^{\beta^{\dagger} \beta}_{ \bd{k}^{\prime} , 
 \bd{k}^{\prime} - \bd{k} } \cdot \bd{e}_{ \bd{k} \lambda} |^2}{2M}
 \delta ( \omega -  E_{\bd{k}^{\prime}} +  E_{\bd{k}^{\prime} - \bd{k}} )
 [ 1 + n (  E_{\bd{k}^{\prime}} )]  n ( E_{\bd{k}^{\prime} - \bd{k}} )
 \nonumber
 \\
&   \hspace{-20mm}  +
 \frac{
  | \tilde{\bd{\Gamma}}^{\beta^{\dagger} \beta^{\dagger}}_{ \bd{k}^{\prime} , 
 \bd{k} - \bd{k}^{\prime}  } \cdot \bd{e}_{ \bd{k} \lambda} |^2}{2M}
 \delta ( \omega -  E_{\bd{k}^{\prime}} - E_{  \bd{k} - \bd{k}^{\prime}} )
 [ 1 + n (  E_{\bd{k}^{\prime}} )]  [1 +n ( E_{   \bd{k} - \bd{k}^{\prime} } )]
 \biggr]
-  ( \omega \rightarrow - \omega, \bd{k} \rightarrow - \bd{k} ).
 \label{eq:selfXmatIm}
 \end{align}
\end{widetext}
The attenuation rate $\gamma_{\bd{k} \lambda}$ 
of a phonon with energy $\omega_{ \bd{k} \lambda}$
can then be obtained from the imaginary part of the self-energy
on resonance:
 \begin{equation}
 \gamma_{\bd{k} \lambda} = - \frac{  {\mathrm{Im}} \Sigma_{2}^{\rm pho} ( 
\omega_{\bd{k} \lambda} + i0, \bd{k} ,  \lambda )}{ 2 \omega_{\bd{k} \lambda} }.
\label{eq:att}
 \end{equation}
At zero temperature we obtain from Eq.~(\ref{eq:selfXmatIm}),
\begin{align}
  \gamma_{\bd{k} \lambda}   & =   \frac{\pi}{ 2 \omega_{\bd{k} \lambda} } \frac{1}{N} 
\sum_{ \bd{k}^{\prime}}
 \frac{
  | \tilde{\bd{\Gamma}}^{\beta^{\dagger} \beta^{\dagger}}_{ \bd{k}^{\prime} , 
 \bd{k} - \bd{k}^{\prime}  } \cdot \bd{e}_{ \bd{k} \lambda} |^2}{2M}
 \nonumber
 \\
 & \times
 \delta ( \omega_{\bd{k} \lambda}-  {E}_{\bd{k}^{\prime}} - 
 {E}_{  \bd{k} - \bd{k}^{\prime}} ).
 \label{eq:damp1}
\end{align}
The leading behavior of $\gamma_{\bd{k} \lambda}$ for small $\bd{k}$ can
be obtained analytically. 
In this limit it is sufficient to use the linear approximation 
(\ref{eq:Eksmall}) for the magnon dispersion. 
Moreover, the interaction vertex 
$\tilde{\bd{\Gamma}}^{ \beta^{\dagger} \beta^{\dagger}   }_{ \bd{k} , \bd{k}^{\prime}}$
can be approximated by the leading 
long-wavelength limit given in Eq.~(\ref{eq:Gammabebelong}).
The resulting integration can then be carried out  analytically and we obtain
 \begin{equation}
\gamma_{\bd{k} \lambda}  =
\frac{\pi^2  }{64}  \left( \frac{ \bd{k}^2}{2M} \right)
 \left( \frac{ S^2  c_{\lambda}^2 \bd{k}^2 }{ V_{\mathrm{BZ}} v_x v_y } \right)
 \frac{ I_{\lambda} ( \hat{\bd{k}}  ) }{ \sqrt{ 1 - r_{\bd{k} \lambda}^2}  },
 \label{eq:dampres1}
\end{equation}
where 
 $
 r_{\bd{k} \lambda}  =  v ( \hat{\bd{k}} ) / c_{\lambda}$,
 and the dimensionless function $I_{\lambda} ( \hat{\bd{k}}  )$ is given by
 \begin{widetext}
 \begin{align}
  I_{\lambda} ( \hat{\bd{k}}  ) & = 
\left( \frac{2 s_{\vt}^2}{c_{\vt}^2} + 1 \right)^2  \left( 1 - r_{\bd{k} \lambda}^2 + 
 \frac{3}{8}  r_{\bd{k} \lambda}^4 \right) 
 \left[ 
{\bd f}^{X\beta}_1(\hat{\bd k})\cdot \bd{e}_{\bd{k} \lambda}
\right]^2 
 \nonumber
 \\
 & +  4 s_{\vt}\frac{v(\hat{\bd k})}{c_\lambda}  \left( \frac{2 s_{\vt}^2}{c_{\vt}^2} + 1 \right)  
 \left( 1 -  
 \frac{3}{4}  r_{\bd{k} \lambda}^2 \right) 
 \left[
{\bd f}^{X\beta}_1(\hat{\bd k})
\cdot \bd{e}_{\bd{k} \lambda}
\right]
 \left[
 {\bd f}^{X\beta}_2(\hat{\bd k},\hat{\bd k})
 \cdot \bd{e}_{\bd{k} \lambda}  
 \right]
 \nonumber
 \\
 & + 
 2\biggl(s_{\vt}\frac{v(\hat{\bd k})}{c_\lambda} \biggr)^2
\left\{ 3 \left[
 {\bd f}^{X\beta}_2(\hat{\bd k},\hat{\bd k})
 \cdot \bd{e}_{\bd{k} \lambda} 
\right]^2
+ ( 1 - r^2_{\bd{k} \lambda} )
\left[
 {\bd f}^{X\beta}_2(\hat{\bd k},\hat{\bd k}_\bot)
 \cdot \bd{e}_{\bd{k} \lambda} 
\right]^2
 \right\}\;.
\label{eq:ilambda}
 \end{align}
\end{widetext}
The vector $\hat{\bd{k}}_{\bot}$ in the last line is given  by  $\hat{\bd{k}}_{\bot} =
 - ( v_y / v_x ) \hat{k}_y \bd{e}_x + (v_x/v_y ) \hat{k}_x \bd{e}_y$.
We conclude that  in the regime of small wavevectors
considered here, the ultrasonic attenuation rate is proportional to $\bd{k}^4$,
which shows that in the presence of 
magnon-phonon interactions
the phonons remain well-defined quasiparticles.
Let us point out that in order to obtain the correct $\bd{k}^4$-dependence of the
ultrasonic attenuation rate, it is crucial to  take the renormalization of the
one-phonon two-magnon vertices 
due to magnon-phonon hybridization into account;
see Eqs.~(\ref{eq:Gmp1}--\ref{eq:deltaGamma3}).
Otherwise one would incorrectly find from Eq.~(\ref{eq:selfX}) that
$\gamma_{\bd{k} \lambda}$ reduces to a nonzero constant for $\bd{k} \rightarrow 0$,
implying that, due to the coupling to the magnon system,
the phonons would cease to be well-defined quasiparticles.
In the appendix, we show that the interaction processes
represented by Fig.~\ref{fig:Feynman}(b), which  involve
intermediate states with one magnon and one phonon, 
also lead to the contribution of the order of $\bd{k}^4$
to the ultrasonic attenuation rate.
Fortunately, this contribution is negligible 
in the experimentally relevant regime where the magnon velocities are
small compared with the phonon velocities.

\section{Comparison with experiments}
\label{sec:comparison}

We have measured the longitudinal $c_{22}$- and $c_{33}$-modes
propagating in the plane of the two-dimensional layered
antiferromagnet Cs$_2$CuCl$_4$ down to temperatures of about $50~\mathrm{mK}$.
Experiments as a function of magnetic field were performed
in the cone state for fields $B$ parallel to the crystallographic $a$-axis.
For the experiments presented here, we used a setup which
allows us to perform simultaneously measurements of changes
in the ultrasonic velocity $\Delta c/c_0$ and the relative
attenuation $\Delta \alpha$ as a function of an external
parameter like temperature or magnetic field. 
We employ a pulse-echo method using a phase-sensitive
detection technique.\cite{Luethi94} Our measurements
can be performed in the frequency range of $5$--$500~\mathrm{MHz}$.
The duration of the ultrasonic echo pulse is between $0.1$ and $1~\mu\mathrm s$,
while the repetition rate is chosen such that it matches the available
cooling power in the cryostat. It lies between $100~\mathrm{Hz}$
in the sub-kelvin temperature range, and a few kiloherz at higher temperatures. 
A high-quality single
crystal of the compound Cs$_2$CuCl$_4$ with the size of 
$3.68 \times 3.8 \times 3.65~\mathrm{mm}^3$
was grown from aqueous solutions by an evaporation 
technique.\cite{Krueger10} Two opposite surfaces
normal to the crystallographic $b$-axis and $c$-axis were polished and a pair of piezoelectric thin-film
transducers was glued to these surfaces. 
These geometries correspond to the longitudinal $c_{22}$
and $c_{33}$ acoustic modes, with the wavevector $\bd k$ and polarization $\bd e_{\bd k\lambda}$
parallel to the crystallographic $b$- and $c$-axis, respectively.\\

To compare the experimental results with our theoretical predictions,
we need realistic estimates for the first  and second derivatives
 $\partial J / \partial x$, $\partial J^{\prime} / \partial r$,
$\partial^2 J / \partial x^2$,
and $\partial^2 J^{\prime} / \partial r^2$
of the exchange couplings for Cs$_2$CuCl$_4$. 
Recall that these derivatives
appear in the explicit expressions (\ref{eq:J1vec})  
and (\ref{eq:J2tensa}--\ref{eq:J2tensc})
for the vector $\bd{J}^{(1)}_{\bd{k}}$
and the tensor $\mathbf{J}^{(2)}_{\bd{k}}$, which
in turn enter our theoretical results for the
elastic constants and the ultrasonic attenuation rate.
In principle, the spatial dependence of the exchange couplings
can be determined using \textit{ab initio} 
methods,~\cite{Zhang08,Foyevtsova2010} but
quantitative \textit{ab initio} results for the derivatives of the exchange
couplings of Cs$_2$CuCl$_4$ are currently not available.
We therefore model the spatial dependence of our two relevant exchange
couplings $J ( x )$ and $J^{\prime} ( r )$ by the following
phenomenological expressions:
 \begin{subequations}
 \begin{align}
 J (x)& =J ( b)  e^{- \kappa   ( x -b)/b }\;,
\label{eq:parametrization_J}
 \\ 
 J^{\prime} (r)& =J^{\prime} ( d )   e^{- \kappa^{\prime}   ( r - d)/d }\;,
 \label{eq:parametrization_Jp}
\end{align}
 \end{subequations}
where $b$ and 
$d  = \sqrt{ b^2 + c^2}/2$ 
are the bond lengths at the equilibrium positions,
and the dimensionless quantities
$\kappa$ and $\kappa^{\prime}$ 
give the inverse range of the exchange interaction in units of the
corresponding inverse equilibrium range. 
Since we do not have any \textit{a priori} knowledge about the numerical values
of $\kappa$ and $\kappa^{\prime}$, we
determine these parameters by fitting our
theoretical predictions for the elastic constants to
the experimental results.

According to Eqs.~(\ref{eq:sigmapho0}) and (\ref{eq:Deltac0})
the classical contribution  $(\Delta c_{\lambda})_{0}$
to the shift in the phonon velocities
arising from the motion of the phonons in the classical spin background
is determined by the dimensionless ratio
\begin{align}
\lim_{ | \bd{k} | \rightarrow 0}    \frac{ \Sigma^{\mathrm{pho}}_0 ( \bd{k} , \lambda )    }{  
 \omega^2_{\bd{k} \lambda}   }  
&=-\frac{S^2}{M c_{\lambda}^2 } \left\{ s_\vt^2 \bigl[ J  {\kappa}^2  +
\frac{J^\prime}{2} 
  {\kappa}^{\prime 2}  \cos^4\varphi_0   \bigr]
\right.\notag\\ 
&  \hspace{-20mm} \left.
+c_\vt^2\bigl[J  {\kappa}^2 \cos(Q_xb) +\frac{J^\prime}{2} 
  {\kappa}^{\prime 2}
\cos^4\varphi_0 \cos( Q_xb /2)\bigr]\right\},
\end{align}
for longitudinal phonons in the $x$-direction ($c_{22}$-mode), and by
\begin{align}
\lim_{ | \bd{k} | \rightarrow 0}    \frac{ \Sigma^{\mathrm{pho}}_0 ( \bd{k} , \lambda )    }{  
 \omega^2_{\bd{k} \lambda}   }  
&=-\frac{S^2}{2M c_{\lambda}^2 }J^\prime {\kappa}^{\prime 2}\sin^4\varphi_0\notag\\
& \times\bigl[ s_\vt^2 +c_\vt^2 \cos(Q_xb/2)\bigr],
\end{align}
for longitudinal phonons in the $y$-direction ($c_{33}$-mode),
where  $J = J ( b)$ and $J^{\prime} = J^{\prime} ( d )$.
Moreover, to evaluate the hybridization contribution
$(\Delta c_{\lambda})_1$ to the velocity shift given in Eq.~(\ref{eq:deltac1})
and the ultrasonic attenuation rate given in Eqs.~(\ref{eq:dampres1}, \ref{eq:ilambda})
we need the dimensionless scalar products
 $ {\bd f}^{X\beta}_1({\hat{\bd k}})\cdot {\bd e}_{{\bd k}\lambda}$
and 
$
 {\bd f}^{X\beta}_2(\hat{\bd k},\hat{\bd k})
 \cdot \bd{e}_{\bd{k} \lambda}  $,
where the vector functions  $ {\bd f}^{X\beta}_1({\hat{\bd k}})$ and
$  {\bd f}^{X\beta}_2({\hat{\bd k}} ,{\hat{\bd k}}^{\prime})$
are defined in Eqs.~(\ref{eq:f1xb}) and (\ref{eq:f2xb}).
To explain the data for the low-temperature measurements of the
 longitudinal c$_{22}$ and c$_{33}$ phonon modes, we set
 ${\hat{\bd k}}=(\hat k_x,\hat k_y)={\bd e}_{{\bd k}\lambda}$. In this case the relevant scalar products simplify to
\begin{align}
 {\bd f}^{X\beta}_1( \hat{\bd k} )\cdot\hat{\bd k}&=-\frac{2i}{h_c}\Bigl\{
\notag\\&\hspace{-10mm}
\hat k_x^2 \bigr[J\kappa (1-\cos(Q_xb))
\notag\\&\hspace{-6mm}
+2J'\kappa' \cos^2\varphi_0(1-\cos(Q_xb/2))\bigl]
\notag\\&\hspace{-10mm}
+\hat k_y^2 2 J' \kappa'\sin^2\varphi_0 (1-\cos(Q_xb/2))\Bigr\}\;,\\
 {\bd f}^{X\beta}_2({\hat{\bd k}} ,{\hat{\bd k}})\cdot {\hat{\bd k}} &
=\frac{i}{v({\hat{\bd k}})}\Bigl\{
\hat k_x^3b\bigl[J\kappa \sin(Q_xb)\notag\\&\hspace{-10mm}
+J'\kappa' \sin(Q_xb/2)\cos^2\varphi\bigr]
\notag\\&\hspace{-14mm}
+3\hat k_x\hat k_y^2 c J'\kappa' \sin\varphi_0\cos\varphi_0\sin(Q_xb/2)\Bigr\}\;,\\
 {\bd f}^{X\beta}_2({\hat{\bd k}} ,\hat{\bd k}_{\bot})\cdot {\hat{\bd k}} &=\frac{i \hat k_x}{v({\hat{\bd k}})}\Bigl\{ 
\notag\\&\hspace{-10mm}
\frac{v_y}{v_x} \hat k_y^2 b\bigl[ J\kappa \sin(Q_xb)
+J'\kappa' \cos^2\varphi_0 \sin(Q_xb/2)\bigr]
\notag\\&\hspace{-10mm}
+\bigl[\frac{v_y}{v_x} \hat k_y^2-\frac{v_x}{v_y}\hat k_x(\hat k_x+\hat k_y)\bigr]
\notag\\&\hspace{-10mm}\times
 c J'\kappa'\sin\varphi_0\cos\varphi_0 \sin(Q_xb/2)\Bigr\}\;,
\end{align}
where we have used the fact that the wavevector $\bd{Q}$ of the spiral has 
only an $x$-component. 

\begin{figure}[tb]\centering
  \includegraphics[width=0.48\textwidth]{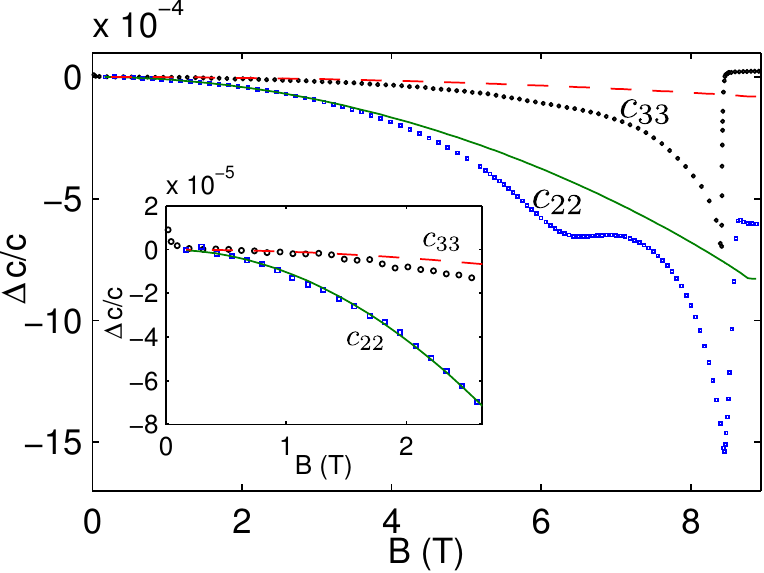}
 \caption{(Color online) 
Measured velocity shifts of the longitudinal 
$c_{22}$-phonon mode (squares) and the $c_{33}$-mode  (circles) 
of Cs$_2$CuCl$_4$ as a function of  the magnetic induction $B$ taken
at  $T= 52~\text{mK}$ ($c_{22}$-mode) and $T= 48~\text{mK}$ ($c_{33}$-mode).
The solid line is a fit of the data for the
$c_{22}$-mode
to our theoretical results at zero temperature
given in Eqs.~(\ref{eq:sigmapho0}, \ref{eq:Deltac0}, \ref{eq:deltac1}, \ref{eq:deltacfull}).
From this fit we obtain for our phenomenological parameters $\kappa$ and $\kappa^{\prime}$
introduced in Eqs.~(\ref{eq:parametrization_J}) and (\ref{eq:parametrization_Jp})
the values
$\kappa \approx 15$ and $\kappa^{\prime} \approx 51$.
Using the same values for the $c_{33}$-mode, we obtain the theoretical
prediction represented  by the dashed line.
From the inset it is obvious that 
in the weak-field regime
$B \lesssim 2.5 $ Tesla where our spin-wave approach should be most accurate
we obtain excellent agreement between theory and experiment.
The anomalies near the critical field $B_c \approx 8.5 $ Tesla 
might have a different physical origin than magnon excitations, so that
these features cannot be explained within our spin-wave approach.
}
\label{fig:fittri}
  \end{figure}

Setting ${\hat{\bd k}}=(1,0)$ for the $c_{22}$-mode and ${\hat{\bd k}}=(0,1)$ for
the $c_{33}$-mode, we can explicitly 
evaluate the terms in Eq.~(\ref{eq:deltacfull})
and calculate the velocity shifts.
In Fig.~\ref{fig:fittri} we compare 
our theoretical results for the magnetic-field dependence of the velocity shifts
of the longitudinal phonon modes with our experimental data
at low temperature. The data for the $c_{22}$-mode was
obtained at $T=52~\mathrm{mK}$, while the data for the $c_{33}$-mode was
taken at $T=48~\mathrm{mK}$.
The parameters $\kappa$ and $\kappa^{\prime}$
were determined by fitting our theoretical results for the
$c_{22}$-mode to the data. 
The resulting values $\kappa \approx 15$ and
$\kappa^{\prime} \approx 51$ are then inserted
back into our expression for the $c_{33}$-mode, so that
our theoretical prediction for the $c_{33}$-mode does not contain any adjustable
parameters.\footnote{Note that the $c_{33}$-phonons do not modulate
the bond lengths of the coupling $J$, so that the associated
velocity shift depends only on $\kappa^\prime$.}
From the inset in Fig.~\ref{fig:fittri} it is obvious that
in the weak-field limit $B \lesssim 2.5 $ Tesla, where our calculations 
of the velocity shifts are expected to be most accurate, 
we obtain excellent agreement between theory and experiment.
The deviations between the experimental data and
our calculations for larger fields signal the breakdown of our  
theoretical approach which
does not take into account higher order fluctuation corrections and 
other types of excitations.
These are likely to
play a role in the vicinity of the critical magnetic field where the
magnetic order vanishes.
Near the critical field for temperatures below $0.3$~K
 the ultrasonic attenuation exhibits a double peak structure 
which will be discussed in a separate publication.

\begin{figure}[tb]
\centering
  \includegraphics[width=0.48\textwidth]{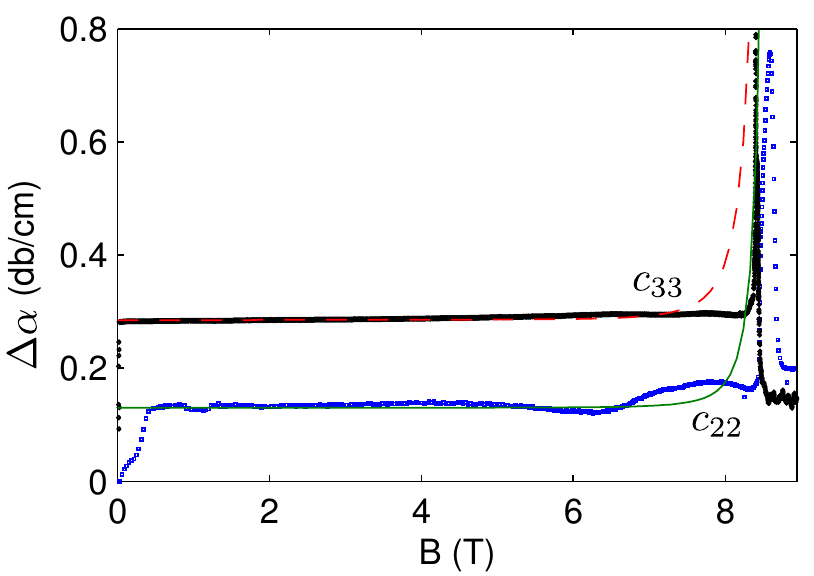}
 \caption{(Color online) 
Experimental results for the relative
ultrasonic attenuation $\Delta \alpha$ in Cs$_2$CuCl$_4$ of the 
longitudinal $c_{22}$-phonon mode (squares) and the $c_{33}$-mode (circles) taken
at  $T= 52~\text{mK}$ ($c_{22}$-mode) and $T= 48~\text{mK}$ ($c_{33}$-mode).
The solid and the dashed lines represent  
our corresponding theoretical predictions
given in Eqs.~(\ref{eq:dampres1}) and (\ref{eq:ilambda}).
The attenuation rates have been obtained
using the same values $\kappa = 15$ and $\kappa^{\prime} = 51$
for the fit parameters as in Fig.~\ref{fig:fittri}.
The  constant offset in the attenuation rates is due to the fact
that with our experimental technique we can only measure
relative attenuation rates.
Note that for very small fields the data approaches $\Delta \alpha=0$.
}
\label{fig:attenuation}
  \end{figure}

Having fixed the fit parameters $\kappa$ and $\kappa^{\prime}$
from the velocity shifts, our theoretical result
for the ultrasonic attenuation rate given in 
Eqs.~(\ref{eq:dampres1}) and (\ref{eq:ilambda}) does not contain
any adjustable parameters.
In Fig.~\ref{fig:attenuation} we compare the results of our calculations
for the attenuation rate of the $c_{22}$ and $c_{33}$ phonon modes
with the experimental data of the relative attenuation $\Delta \alpha$.
For fields in the range $h \lesssim 0.8 h_c$ the 
measured attenuation rates are rather small, while one observes a strong enhancement
for $h \rightarrow h_c$. This indicates that, at least for small magnetic
fields, other effects leading to the large attenuation background
are not field dependent. The overall shape of the
data is reproduced by our theoretical curves rather well, although our spin-wave calculation
cannot reproduce the shoulder just below the critical field. 
Note that according to
Eqs.~(\ref{eq:dampres1}) and (\ref{eq:ilambda})
the attenuation rate for fields
in the vicinity of the critical field  can  be approximated by
\begin{equation}
\gamma_{\bd{k} \lambda}  \approx
\frac{\pi^2  }{64}  \left( \frac{ \bd{k}^2}{2M} \right)
 \left( \frac{ S^2  c_{\lambda}^2 \bd{k}^2 }{ V_{\mathrm{BZ}} v_x v_y } \right)
 \frac{  \left[ 
{\bd f}^{X\beta}_1(\hat{\bd k})\cdot \bd{e}_{\bd{k} \lambda}
\right]^2     }{( 1 - h / h_c )^2   },
 \label{eq:dampressmall}
\end{equation}
where we have approximated
$\sin \theta \approx  1$ and
$\cos \theta \approx \sqrt{ 2 ( 1 - h / h_c )}$ for $h$ close to $h_c$, and have
used the fact that in Cs$_2$CuCl$_4$ 
the ratio
$ r_{\bd{k} \lambda} = v ( \hat{\bd{k}} ) / c_{\lambda}$
is small compared with unity. 
One should keep in mind, however, that in the vicinity of $h_c$
higher order fluctuation corrections which we have neglected within our
spin-wave expansions are likely to become important, so that
we do not expect that Eq.~(\ref{eq:dampressmall}) is quantitatively accurate
very close to $h_c$. Nevertheless, 
from Fig.~\ref{fig:attenuation} it is clear that
the factor of $(1 - h / h_c )^{-2}$
in  Eq.~(\ref{eq:dampressmall}) gives a satisfactory
description of the strong enhancement
of the attenuation rate in the vicinity of the critical field.

\section{Summary and Conclusions}
 \label{sec:summary}

In summary, we have
presented both theoretical and experimental results for the
magnetic-field dependence of the elastic constants
and the ultrasonic attenuation rate in the cone state
of the frustrated quantum antiferromagnet
Cs$_2$CuCl$_4$.
Our calculations are based on the assumption that
two-dimensional magnons are well defined excitations in this material, 
so that we may use the $1/S$ spin-wave expansion in combination with
an expansion in powers of phonon operators. 
Using a simple phenomenological 
parametrization of the
spatial variations of the 
exchange couplings involving two adjustable parameters,
our theoretical results for the 
magnetic-field dependence of the elastic constants
agree quite well with our 
experimental measurements, in particular in the weak-field regime
$B \lesssim 2.5$ Tesla where our
perturbative approach is expected to be most accurate.
Our theoretical results for the ultrasonic attenuation rate
reproduce the strong enhancement close to the
critical magnetic field, although our approach is
likely to break down in the vicinity of the quantum critical point.

The fact that the ultrasonic attenuation rate
is proportional to $\bd{k}^4$ implies that the coupling to the
magnon excitations does not destroy the quasiparticle character of the
long-wavelength acoustic phonons. 
To arrive at this result, it was essential to work out the
relevant magnon-phonon matrix elements with great care,
taking into account that 
due to magnon-phonon hybridization in the cone state
the cubic part the pure magnon Hamiltonian
indirectly contributes to the effective one-phonon two-magnon interaction.
Up to now the predicted $\bd k^4$-dependence of the attenuation rate on
the phonon momentum could not be detected experimentally.
A possible reason is that the magnetic contribution is small compared
with the background, which does not depend on the magnetic field. In the
vicinity of the critical field, the magnetic contribution becomes large,
but it is questionable whether our calculation is still valid in this regime.

We have intentionally restricted the calculations presented in this paper
to the magnetically ordered cone state of Cs$_2$CuCl$_4$,
because in this case our spin-wave approach is well justified.
The investigation of magnon-phonon interactions
in the immediate vicinity of the quantum critical point and in the other phases of
Cs$_2$CuCl$_4$ such as the spin-liquid phase or the ferromagnetic phase
is left for future work.
Since magnons in frustrated magnets can have
anomalous properties such as strong damping,\cite{Chernyshev06,Chernyshev09a}
due to magnon-magnon interactions,
it should be interesting to take a closer look at these interaction processes
between magnons
in the cone state, especially in the vicinity of the quantum critical point.

\section*{ACKNOWLEDGMENTS}
We thank K. Foyevtsova, M. Malard, and R. Valent{\'{\i}}  for discussions.
This work was financially supported by the DFG via SFB/TRR 49 and by the DAAD via the PROBRAL program.

\appendix

\renewcommand{\theequation}{A.\arabic{equation}}
\renewcommand{\thesubsection}{\arabic{subsection}}
\section*{APPENDIX: 
PROCESSES INVOLVING ONE PHONON AND ONE MAGNON IN THE INTERMEDIATE STATE}


In this appendix, we show that 
in  Cs$_2$CuCl$_4$ 
where the phonon velocities are large compared with the magnon velocities
the ultrasonic attenuation rate
due to processes involving one phonon and one magnon in the
intermediate state shown in  Fig.~\ref{fig:Feynman}(b)
is small compared with the contribution 
from processes with two magnons in the intermediate state
shown in Fig.~\ref{fig:Feynman}(a).


Consider the part $H^{\mathrm{2pho}}_{\mathrm{1mag}}$
of the magnon-phonon interaction involving
two powers of the phonon operators and one power
of the  magnon operators.
It can be obtained from $H^{\mathrm{npho}}_{\mathrm{spin}}$
given in  Eq.~(\ref{eq:Usp})
by setting $n=2$ and expanding the spin operators
to first order in the Holstein--Primakoff  
magnon operators.
The Fourier transform of the
relevant phonon part $U^{(2)}_{\bd{k} , \bd{k}^{\prime}}$
is given in Eq.~(\ref{eq:U2mat}). 
We obtain
 \begin{align}
H^{\mathrm{2pho}}_{\mathrm{1mag}} & = \frac{1}{2! \sqrt{N} } \sum_{ \bd{k} , \bd{k}^{\prime}} 
 \sum_{ \lambda \lambda^{\prime}} 
 \Bigl[
 \Gamma^{XX {b}^{\dagger}}_{ \bd{k} \lambda,  \bd{k}^{\prime}  \lambda^{\prime}}
 X_{ \bd{k} \lambda} X_{ \bd{k}^{\prime} \lambda^{\prime}} b^{\dagger}_{ 
 \bd{k} + \bd{k}^{\prime}}
 \nonumber
 \\
 & \hspace{10mm} +\Gamma^{XXb}_{ \bd{k} \lambda,  
\bd{k}^{\prime}  \lambda^{\prime}}
 X_{ -\bd{k} \lambda} X_{ -\bd{k}^{\prime} \lambda^{\prime}} b_{ 
 \bd{k} + \bd{k}^{\prime}} \Bigr],
 \label{eq:H2p1m}
\end{align}
where the vertices are 
 \begin{align}
 \Gamma^{XX {b}^{\dagger} }_{ \bd{k} \lambda, \bd{k}^{\prime}  \lambda^{\prime}}
 & = \bd{e}_{\bd{k} \lambda}^{\dagger} 
\mathbf{ \Gamma}^{XX {b}^{\dagger} }_{ \bd{k} \bd{k}^{\prime}}
 \bd{e}_{\bd{k}^{\prime} \lambda^{\prime}},
 \\
\Gamma^{XXb}_{ \bd{k} \lambda,  \bd{k}^{\prime} \lambda^{\prime}}
 & = \bd{e}_{\bd{k} \lambda}^{\dagger} 
\mathbf{ \Gamma}^{XXb}_{ \bd{k} \bd{k}^{\prime}}
 \bd{e}_{\bd{k}^{\prime} \lambda^{\prime}},
 \end{align}
with $\mathbf{ \Gamma}^{XX \bar{b}}_{ \bd{k} \bd{k}^{\prime}}$
and $\mathbf{ \Gamma}^{XXb}_{ \bd{k} \bd{k}^{\prime}}$
given by  the following matrices in the direction labels,
\begin{align}
 \mathbf{ \Gamma}^{XX {b}^{\dagger}}_{ \bd{k} \bd{k}^{\prime}} & =
 \frac{i}{4} ( 2 S)^{3/2} c_{\theta} \Bigl[
\mathbf{J}^{(2-)}_{ \bd{k} + \bd{k}^{\prime} , \bd{Q}}
 - \mathbf{J}^{(2-)}_{ \bd{k}  , \bd{Q}}
 - \mathbf{J}^{(2-)}_{ \bd{k}^{\prime}  , \bd{Q}} 
 \nonumber
 \\
 &
 - s_{\theta} 
 \bigl(  \mathbf{J}^{(2+)}_{ \bd{k} + \bd{k}^{\prime} , \bd{Q}}
 - \mathbf{J}^{(2+)}_{ \bd{k}  , \bd{Q}}
 - \mathbf{J}^{(2+)}_{ \bd{k}^{\prime}  , \bd{Q}} 
  + \mathbf{J}^{(2+)}_{ 0  , \bd{Q}} 
 \bigr) \Bigr]
 \nonumber
 \\
 & =
 -  \mathbf{ \Gamma}^{XX {b}}_{ \bd{k} \bd{k}^{\prime}},
 \end{align}
where the tensors  $\mathbf{J}^{(2\pm)}_{ \bd{k}  , \bd{Q}}$ were introduced in Eqs.~(\ref{eq:J2m}, \ref{eq:J2p}).
From the symmetries in \Ref{eq:symmJ2} it follows that for small $\bd{k}$ and $\bd{k}^{\prime}$ the vertices 
$\mathbf{ \Gamma}^{XX \bar{b}}_{ \bd{k} \bd{k}^{\prime}}$
and $\mathbf{ \Gamma}^{XXb}_{ \bd{k} \bd{k}^{\prime}}$ vanish quadratically
in $\bd{k}$ and $\bd{k}^{\prime}$.
Transforming to the Bogoliubov basis, our Hamiltonian (\ref{eq:H2p1m})
assumes the form
 \begin{align}
H^{2\mathrm{pho}}_{\mathrm{1mag}} & = \frac{1}{2! \sqrt{N} } \sum_{ \bd{k} , \bd{k}^{\prime}} 
 \sum_{ \lambda \lambda^{\prime}} 
 \Bigl[
 \Gamma^{XX \bar{\beta} }_{ \bd{k} \lambda,  \bd{k}^{\prime} \lambda^{\prime}}
 X_{ \bd{k} \lambda} X_{ \bd{k}^{\prime} \lambda^{\prime}} \beta^{\dagger}_{ 
 \bd{k} + \bd{k}^{\prime}}
 \nonumber
 \\
 & \hspace{10mm} +\Gamma^{XX \beta}_{ \bd{k} \lambda, \bd{k}^{\prime} \lambda^{\prime}}
 X_{ -\bd{k} \lambda} X_{ -\bd{k}^{\prime} \lambda^{\prime}} \beta_{ 
 \bd{k} + \bd{k}^{\prime}} \Bigr],
 \label{eq:H2p1mbog}
\end{align}
with
 \begin{align}
\mathbf{\Gamma}^{XX \bar{\beta}}_{ \bd{k} \bd{k}^{\prime} }
& =  u_{ \bd{k} + \bd{k}^{\prime}}
\mathbf{\Gamma}^{XX \bar{b}}_{ \bd{k} \bd{k}^{\prime} }
- v_{ \bd{k} + \bd{k}^{\prime}}
\mathbf{\Gamma}^{XXb}_{ - \bd{k}, - \bd{k}^{\prime} },
 \label{eq:GammaXXbeta1}
 \\
\mathbf{\Gamma}^{XX \beta }_{ \bd{k} \bd{k}^{\prime} }
& =  u_{ \bd{k} + \bd{k}^{\prime}}
\mathbf{\Gamma}^{XXb}_{ \bd{k} \bd{k}^{\prime} }
- v_{ \bd{k} + \bd{k}^{\prime}}
\mathbf{\Gamma}^{XX \bar{b}}_{ - \bd{k}, - \bd{k}^{\prime} }.
 \label{eq:GammaXXbeta2}
 \end{align}


To second order
in the two-phonon one-magnon vertex, we obtain for the phonon self-energy
 \begin{align}
 \Sigma^{\rm pho}_3 ( K , \lambda )   & =  
\frac{T}{ N}\! \sum_{ K^{\prime}  \lambda^{\prime}}\!
 \frac{  | \Gamma^{XX {\beta} }_{  \bd{k} \lambda , \bd{k}^{\prime}  \lambda^{\prime}} |^2 }{M^2}  G^{\rm pho} ( K^{\prime} \lambda^{\prime} ) G_{\rm mag} ( K^{\prime} \!+\! K )
 \nonumber
 \\
 & 
+ ( K \rightarrow - K ),
 \end{align}
where $\Gamma^{XX \beta}_{ \bd{k} \lambda,  \bd{k}^{\prime} \lambda^{\prime}}
  =  \bd{e}_{\bd{k} \lambda}^{\dagger} 
\mathbf{ \Gamma}^{XX \beta}_{ \bd{k} \bd{k}^{\prime}}
 \bd{e}_{\bd{k}^{\prime} \lambda^{\prime}}$.
The Matsubara sums can now be carried, out and we obtain
\begin{align}
  \Sigma^{\rm pho}_3 ( K , \lambda )   & = 
  \frac{1}{ N} \sum_{ \bd{k}^{\prime}   \lambda^{\prime}}
\frac{| \Gamma^{XX {\beta} }_{ \bd{k} \lambda , - \bd{k}^{\prime}  
 \lambda^{\prime} }  |^2}{2   \omega_{ \bd{k}^{\prime} \lambda^{\prime}}  M^2}
 \biggl[  
\frac{ n ( \omega_{\bd{k}^{\prime} \lambda^{\prime}} ) -   
n ( E_{   \bd{k}  - \bd{k}^{\prime}} )}{ i \omega 
 + \omega_{\bd{k}^{\prime} \lambda} - E_{ \bd{k} - \bd{k}^{\prime} } }
 \nonumber
 \\
 &  \hspace{-15mm} +
 \frac{ n ( \omega_{\bd{k}^{\prime} \lambda^{\prime}} ) 
+   n ( E_{  \bd{k} -   \bd{k}^{\prime}  } ) + 1}{  i \omega 
 - \omega_{\bd{k}^{\prime} \lambda^{\prime}} - E_{  \bd{k} -   \bd{k}^{\prime}   } } \biggr]
+ ( K \rightarrow - K ),
 \label{eq:selfXmat}
 \end{align}
At zero temperature this yields for the ultrasonic attenuation rate
\begin{align}
  \gamma_{\bd{k} \lambda}   & =   \frac{\pi}{ 2 \omega_{\bd{k} \lambda} } \frac{1}{N} 
\sum_{ \bd{k}^{\prime} \lambda^{\prime}}
 \frac{\bigl| \Gamma^{XX {\beta} }_{\bd{k} \lambda , - \bd{k}^{\prime}  
 \lambda^{\prime} }  \bigr|^2}{2   \omega_{ \bd{k}^{\prime} \lambda^{\prime}}  M^2}
 \delta ( \omega_{\bd{k} \lambda}-  {\omega}_{\bd{k}^{\prime} \lambda^{\prime} } - 
 {E}_{  \bd{k} -   \bd{k}^{\prime} } ).
 \label{eq:damp2p1m}
\end{align}
In the long-wavelength limit we obtain for the relevant matrix element 
of the interaction vertex
 \begin{align}
  \bd{\Gamma}^{XX {\beta} }_{\bd{k}  , - \bd{k}^{\prime}   } 
 & =  - \frac{i}{4} ( 2 S)^{3/2} \Biggl\{
 \nonumber
 \\
 & \hspace{-10mm}
 s_{\theta} \sqrt{ \frac{\epsilon_{ \bd{k} - \bd{k}^{\prime}} }{h_c} }
 ( \bd{k} \cdot \nabla_{\bd{Q}} )  ( \bd{k}^{\prime} \cdot \nabla_{\bd{Q}} ) 
  \left[ \left.  {\mathbf{J}}^{(2)}_{\bd{Q}} \right|_{\bd{Q}=0} -     {\mathbf{J}}^{(2)}_{\bd{Q}} \right]
 \nonumber
 \\
 & \hspace{-10mm}  -
 \frac{c_{\theta}^2}{2} \sqrt{ \frac{h_c}{\epsilon_{ \bd{k} - \bd{k}^{\prime}} } }
[ ( \bd{k} - \bd{k}^{\prime} ) \cdot \nabla_{\bd{Q}} ]
 ( \bd{k} \cdot \nabla_{\bd{Q}} )  ( \bd{k}^{\prime} \cdot \nabla_{\bd{Q}} ) 
 {\mathbf{J}}^{(2)}_{\bd{Q}}
 \Biggr\}.
\end{align}
To carry out the integration in Eq.~(\ref{eq:damp2p1m}),
we 
use the fact that in the experimentally relevant regime the phonon velocities are much larger than the magnon velocities.
Then the scattering surface defined by 
$\omega_{\bd{k} \lambda} -   {\omega}_{\bd{k}^{\prime} \lambda^{\prime} } - 
 {E}_{  \bd{k} -   \bd{k}^{\prime} } =0$ can be explicitly calculated to leading order
in $v ( \hat{\bd{k}} ) / c_{\lambda}$.
Using circular coordinates we obtain the parametric representation
 \begin{equation} 
 \bd{k}^{\prime} ( \varphi^{\prime} ) = | \bd{k} |  \frac{ c_{\lambda} 
 - u ( \hat{\bd{k}} , \varphi^{\prime} ) }{c_{{\lambda}^{\prime}}} 
\hat{\bd{k}}^{\prime} (\varphi^{\prime} ) ,
 \end{equation}
where
 $
 \hat{\bd{k}}^{\prime} (  \varphi^{\prime} ) =
 \cos \varphi^{\prime} \bd{e}_x +  \sin \varphi^{\prime} \bd{e}_y$,
and
 \begin{equation}
 u ( \hat{\bd{k}} , \varphi^{\prime} ) =
 \sqrt{ v_x^2 ( \hat{k}_x - \cos \varphi^{\prime} )^2 +  
v_y^2 ( \hat{k}_y - \sin \varphi^{\prime} )^2 }.
 \end{equation}
 \begin{widetext}
To leading order in $ v ( \hat{\bd{k}} ) / c_{\lambda}$ we obtain
 \begin{align}
 \gamma_{\bd{k} \lambda} & =  \frac{\pi S^3 }{4} 
 \left( \frac{ \bd{k}^2}{2M} \right)  \left( \frac{  \bd{k}^2 }{ V_{\mathrm{BZ}}}\right)
 \sum_{\lambda^{\prime}}  \left( \frac{h_c}{ M c_{\lambda^{\prime}}^2} \right)
 \left( \frac{c_{\lambda}}{  c_{\lambda^{\prime}} } \right)^2
\int_{0}^{2 \pi}
 d \varphi^{\prime}  \frac{  u ( \hat{\bd{k}} , \varphi^{\prime} )}{ c_{\lambda^{\prime}}}
 | \bd{e}_{\bd{k} \lambda}^{\dagger} \mathbf{F}^{XX \beta} ( \hat{\bd{k}} , \varphi^{\prime} ) 
 \bd{e}_{\bd{k}^{\prime} \lambda^{\prime}} |^2 ,
 \label{eq:dampres2}
 \end{align}
where the dimensionless matrix element is given by
 \begin{align}
  \mathbf{F}^{XX \beta } ( \hat{\bd{k}}  , \varphi^{\prime}   )
 &   = 
 \frac{s_{\theta}}{h_c} 
( \hat{\bd{k}} \cdot \nabla_{\bd{Q}} ) 
 ( \hat{\bd{k}}^{\prime} ( \varphi^{\prime} ) \cdot \nabla_{\bd{Q}} )  
  \Bigl[ \left.  {\mathbf{J}}^{(2)}_{\bd{Q}} \right|_{\bd{Q}=0} -     {\mathbf{J}}^{(2)}_{\bd{Q}} \Bigr]
 \nonumber
 \\
 &   -
 \frac{c_{\theta}^2}{2  u ( \hat{\bd{k}} , \varphi^{\prime} )
  }
[ ( \hat{\bd{k}}  - \frac{ c_{\lambda}}{c_{\lambda^{\prime}}}
\hat{\bd{k}}^{\prime} ( \varphi^{\prime} ) ) \cdot \nabla_{\bd{Q}} ]
 ( \hat{\bd{k}} \cdot \nabla_{\bd{Q}} )  ( \hat{\bd{k}}^{\prime} ( \varphi^{\prime} ) 
\cdot \nabla_{\bd{Q}} )  {\mathbf{J}}^{(2)}_{\bd{Q}}.
\end{align}
 \end{widetext}
Comparing 
Eq.~(\ref{eq:dampres2}) with the corresponding damping rate 
(\ref{eq:dampres1}) due to  one-phonon two-magnon processes, we see
that in the experimentally relevant regime where the magnon velocities are
small compared with the phonon velocities, the processes with two magnons in the
intermediate states are dominant, although  both processes yield
contributions to the ultrasonic attenuation rate which are proportional to
$\bd{k}^4$ at long wavelengths.

\end{document}